\documentclass[aps,prx,twocolumn,superscriptaddress]{revtex4}
\usepackage{amsmath, amsthm}
\usepackage{keyval}
\usepackage{graphicx,latexsym}
\usepackage{dcolumn}
\usepackage{bm}
\usepackage{color}
\usepackage{ulem} % strikeout \sout
\usepackage{hyperref}
\usepackage{xr}

\definecolor{magenta}{rgb}{1,0,1}
\usepackage{color}

\hypersetup{
    colorlinks=true, 
    linktoc=all,   
    linkcolor=blue,  
}

\begin{document}
\title{Interlayer couplings in cuprates:  structural origins, analytical forms, and structural estimators}
\author{Zheting Jin}
\affiliation{Department of Applied Physics, Yale University, New Haven, Connecticut 06520, USA}
\author{Sohrab Ismail-Beigi}
\affiliation{Department of Applied Physics, Yale University, New Haven, Connecticut 06520, USA}
\affiliation{Department of Physics, Yale University, New Haven, Connecticut 06520, USA}
\affiliation{Department of Mechanical Engineering and Materials Science, Yale University, New Haven, Connecticut 06520, USA}
\date{\today}
\begin{abstract}
We quantitatively identify the multiple distinct microscopic mechanisms contributing to effective interlayer couplings (EICs) by performing first-principle calculations for two prototype superconducting cuprate families, pristine and doped Bi$_2$Sr$_2$CaCuO$_2$O$_{8+x}$ and Pr$_{x}$Y$_{1-x}$Ba$_2$Cu$_3$O$_7$.  The major mechanisms are mediated by interlayer oxygen $p_{\sigma}$-$p_{\sigma}$ and $p_z$-$p_z$ hoppings as well as interlayer copper $d_{z^2}$-oxygen $p_{\sigma}$ hoppings.  Furthermore, we show how EICs are closely related to structural distortions such as layer bucklings and bond length changes.  This allows us to provide analytical formulae that permit direct estimation of the key interatomic hoppings and the EICs based only on the crystal structure.  Finally, we benchmark our method on YBa$_2$Cu$_3$O$_7$ to estimate the strength and anisotropy of the EIC.   
\end{abstract}
\maketitle
\section{Introduction}

Cuprates have a quasi-two-dimensional structure, consisting of superconducting copper-oxide layers separated by charge reservoir layers.  While the high-temperature superconductivity was long believed to live on the copper-oxide layers, more and more recent experiments revealed that the out-of-plane physics also plays a crucial role in cuprates, affecting three-dimensional charge orders \cite{gerber2015three, chang2016magnetic, jang2016ideal, bluschke2018stabilization, jang2022characterization}, charge-transfer gap sizes \cite{ruan2016relationship, wang2023correlating}, and critical temperatures ($T_c$) \cite{andersen2007superconducting, ruan2016relationship, wang2023correlating}.  In particular, changing the number of neighboring CuO$_2$ layers is one of the most efficient ways to increase $T_c$ \cite{chen1999dependence, khandka2007effect, iyo2007t}.  For optimal doping, $T_c$ of multilayer cuprates are generally higher than single-layer cuprates.  Understanding effective interlayer couplings between copper-oxide layers is fundamental to unraveling the mechanism of high-temperature superconductivity and designing novel materials to increase $T_c$.

The electronic spectra of multilayer cuprates have been intensively measured by angle-resolved photoemission spectroscopy (ARPES), where the EICs between different CuO$_2$ planes cause band splitting \cite{damascelli2003angle}.  Many studies have attempted to reveal the strength of EICs by fitting the band splitting into simple tight-binding Hamiltonians, where one can reproduce the experimental band dispersion based on the fittings \cite{chakravarty1993interlayer, ideta2010enhanced, ideta2021hybridization, luo2023electronic}.  However, the theoretical interpretation has limited prediction power due to the phenomenological nature of the fitting theory.  In addition, due to the anisotropy of the band splitting, one has to decide on which formula to fit into.  To date, the most prominent empirical model for cuprate superconductors is a Cu-only model (using $d_{x^2-y^2}$ ($d_{x^2}$) symmetry local orbitals) where the EIC between two Cu $d_{x^2}$ orbitals is 
\begin{equation}
\label{equ:old}
t(k) = t_0 + t_1(\text{cos}(k_xa) – \text{cos}(k_ya))^2/4\,,
\end{equation}
where $t_0 $ and $t_1$ are the two fitting parameters.  The explicit contributions from oxygen atoms and other out-of-plane orbitals are missing in this effective one-band model, and its microscopic basis is not obvious.  In addition, the fit parameters $t_0 $ and $t_1$ are material-specific which limits the transferability of this model.  In practice, this formula fails to reproduce key spectral properties in trilayer BSCCO in a recent ARPES study \cite{luo2023electronic}.  One possible improvement for trilayer BSCCO is to include the hopping between the two distant outer CuO$_2$ planes using the same form as Eq. \eqref{equ:old}, but the longer-ranged hopping parameters resulting from such a fit turn out to be much larger than the hoppings between adjacent CuO$_2$ planes \cite{luo2023electronic} which is counter-intuitive.  

% Another possible thing to talk about in the intro is the importance of considering structural distortion

These shortcomings leave us with some questions to answer: 1. What are the dominant microscopic mechanisms of the effective interlayer couplings between Cu $d_{x^2}$ orbitals?  2. What are the physical origins of the anisotropy in $k$-space and are there better analytical formulae describing the anisotropy?  3. How should one estimate the EIC without comprehensive theoretical calculations or high-quality spectral experiments?  To address these questions, a microscopic understanding and a more general predictive theory are needed.

We use density functional theory (DFT) to study the microscopic mechanisms contributing to the EICs between Cu $d_{x^2}$ orbitals on prototype bilayer cuprates.  These mechanisms are then classified by the intermediate hoppings that connect orbitals from different layers.  The major interlayer connections include interlayer oxygen $p_{\sigma}$-$p_{\sigma}$ and $p_z$-$p_z$ hoppings and interlayer copper $d_{z^2}$-oxygen $p_{\sigma}$ hopping.  These participating orbitals are then connected to the $d_{x^2}$ orbitals via in-plane hoppings, bridging $d_{x^2}$ orbitals from different layers.  Next, we find analytical formulae that directly relate the participating hopping strengths to corresponding structural properties such as layer distortions and bond lengths.  These formulae show invariance across different cuprate materials: this transferability allows one to use them directly in many cuprates.  We then derive formulae that can directly estimate the EICs based only on the crystal structure of the material.  We test our theory on YBCO7 by estimating the anisotropic EICs which match well with the directly computed DFT results.  The main terminologies and abbreviations used in this article are listed in Table~\ref{tab:term}.  
%%%%%%%%%%%%% TABLE %%%%%%%%%%%%%%%%%%%%%%%%%%%%%%%%%%%
\begin{table} 
\caption{
Terminologies and abbreviations.  In-plane refers to a single CuO$_2$ layer, and interlayer refers to processes between two adjacent CuO$_2$ layers.
}
\begin{ruledtabular} 
\begin{tabular} { c c}
Name & Meaning\\
\hline 
undoped BSCCO & Bi$_2$Sr$_2$CaCu$_2$O$_8$ \\
hole-doped BSCCO & Bi$_2$Sr$_2$CaCu$_2$O$_{8.25}$ \\
YBCO7 & YBa$_2$Cu$_3$O$_7$ \\
Y-site PrYBCO7 & Pr$_{0.5}$Y$_{0.5}$Ba$_2$Cu$_3$O$_7$ \\
Ba-site PrYBCO7 & Pr$_{0.5}$YBa$_{1.5}$Cu$_3$O$_7$ \\
$p_{\sigma}$ & Oxygen $p$-orbital pointing towards Cu\\
$t_{ab}^\perp$ & Interlayer hopping from orbital $a$ to $b$ \\
$t_{ab}^\parallel$ & In-plane hopping from orbital $a$ to $b$ \\
$\Delta_{tot}$ & Total band splitting caused by EIC\\
$\Delta_{ab\perp}$ & EIC band splitting mediated by $t_{ab}^\perp$
\end{tabular} 
\end{ruledtabular} 
\label{tab:term}
\end{table}

%%%%%%%%%%%%%%%%%%%%%%%%%%%%%%%%%%%%%%%%%%%%%%%%%%%%%%%

\section{Band splittings}
We begin with DFT calculations on undoped/hole-doped BSCCO, YBCO7, Y- and Ba-site PrYBCO7 using the Vienna ab initio simulation package (VASP) software \cite{kresse1996efficiency, kresse1996efficient}.  The generalized-gradient-approximation (GGA) with the semilocal Perdew–Burke–Ernzerhof (PBE) functional \cite{anisimov1991band, perdew1996generalized} is used in all of our calculations.  Additional DFT+$U$ correction with $U=4$ eV was also applied to Cu $d$-orbitals to correct the self-interaction errors (SIE) \cite{perdew1981self} following previous theoretical works \cite{yelpo2021electronic, wang2006oxidation, deng2019higher, jin2023afm}. Note that the choice of functional, the choice of $U$-values in DFT+$U$, and different magnetic orders have a negligible impact ($<4\%$) on the hopping strengths we focus on in this work. Detailed comparison of DFT calculations with different functionals such as strongly-constrained-and-appropriately-normed (SCAN) and SCAN+$U$ functionals are tabulated in Appendix~\ref{app:functional}.  
Other computational details as well as structural models are found in Appendix~\ref{app:DFT}. 

To extract hopping parameters, we compute the tight-binding Kohn-Sham Hamiltonian in the maximally localized Wannier basis \cite{marzari1997maximally,souza_maximally_2001} extracted from our DFT calculations using the Wannier90 software \cite{pizzi2020wannier90}.  The Wannierized orbitals include all Cu $d$-orbitals and O $p$-orbitals.  In addition, for BSCCO, we also include Bi $p$-orbitals because their energies lie close to those of the Cu $d$-orbitals.  Other orbitals localized on Y, Pr, Ba, Ca, Sr, as well as Cu-$p$ and O-$s$ are either occupied semi-core orbitals or high-energy empty orbitals and couple weakly to states near the Fermi energy.  We find that the Wannierized manifold of Cu-$d$, O-$p$, and Bi-$p$ orbitals are sufficiently accurate to describe the DFT band structures from about -3 eV to +3 eV around the Fermi level.  Detailed comparisons of the band structures obtained by DFT calculations and Wannierized tight-binding models are found in Appendix~\ref{app:DFT}. 

%%%%%%%%%%% FIGURE %%%%%%%%%%%%%%%%%%%%%%%%%%%%%%%%%%%
\begin{figure}[t]
\begin{center}
\includegraphics[scale=0.6]{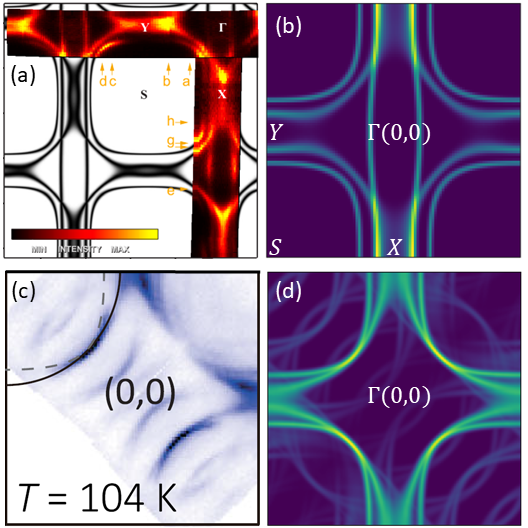}
\end{center}
\caption{
Experimental and theoretical Fermi surfaces.  (a) ARPES-measured Fermi surface of untwinned YBa$_2$Cu$_3$O$_{6.85}$ and the associated tight-binding fit adapted from existing experiments \cite{zabolotnyy2007momentum}.  Note that the origin $\Gamma (0,0)$ is not at the center but at the top right. (b) DFT Fermi surface of YBa$_2$Cu$_3$O$_{7}$ at $k_z=0$, horizontal and vertical axes are $k_x$ and $k_y$ ranging from $-\pi/a$ to $\pi/a$. (c) ARPES-measured Fermi surface of $x=26$\% hole-doped BSCCO measured at $T=104$K, adapted from experiments \cite{he2021superconducting}.  (d) DFT Fermi surface of $x=25$\% hole-doped BSCCO.   
}
\label{fig:Fermi_surface}
\end{figure}
%%%%%%%%%%%%%%%%%%%%%%%%%%%%%%%%%%%%%%%%%%%%%%%%%%%%%%

Fig.~\ref{fig:Fermi_surface} shows the Fermi surfaces of two prototyical bilayer cuprates measured by ARPES \cite{zabolotnyy2007momentum, he2021superconducting} and computed from our Wannierized tight-binding model.  To facilitate comparison to experimental ARPES spectra of the Fermi surface, we unfold bands of the large supercell onto the primitive unit cell Brillouin zone following standard band-unfolding method \cite{ku2010unfolding, brouet2012impact}.  The band unfolding approach is known to qualitatively reproduce the ARPES spectral intensities in various materials \cite{lin2011one, medeiros2014effects, tomic2014unfolding, zhu2018quasiparticle}.  
%\red{How did you compute the Fermi surfaces?  You used band folding?  You need to write a sentence with references or explanation for what is being shown} 
For the theoretical results shown in Fig.~\ref{fig:Fermi_surface}(b) and (d), the Cu atoms are set to be non-magnetic.  This is because computed ARPES spectra in a non-magnetic state of the hole-doped cuprates compare well to ARPES measurements \cite{damascelli2003angle, sobota2021angle, jin2023afm}.  As a result, the theoretical Fermi surface of YBCO7 and hole-doped BSCCO in Fig.~\ref{fig:Fermi_surface}(b) and (d) match well with the ARPES measurements in Fig.~\ref{fig:Fermi_surface}(a) and (c), respectively.

Despite differences in Fermi surface between various cuprates, they do share some key similarities.  First, for bilayer cuprates, there are always two main Fermi arcs with the strongest intensity connecting the nodal (around (0.42$\pi/a$, 0.42$\pi/a$)) and antinodal (around (0.18$\pi/a$, $\pi/a$)) regions.  The band splitting between these two main Fermi arcs reveals the strength of the EIC $\Delta_{tot}$, where the arc closer to the $\Gamma$ point is the antibonding band while the other arc is the bonding bond \cite{feng2001bilayer}.  Second, the band splitting depends strongly on the $k$-point: it is generally larger around the antinodal region than the nodal region.  Based on the close agreement between theory and experiment, we are encouraged to use the DFT results to explore the physical mechanisms controlling the EIC $\Delta_{tot}(k)$.  As a convention in this work, we use the interlayer band splitting $\Delta$ as the proxy representing the strength of the EIC. 

\section{Identifying Microscopic Mechanisms}
\label{sec:zeroing}
%%%%%%%%%%% FIGURE %%%%%%%%%%%%%%%%%%%%%%%%%%%%%%%%%%%
\begin{figure}[t]
\begin{center}
\includegraphics[scale=0.37]{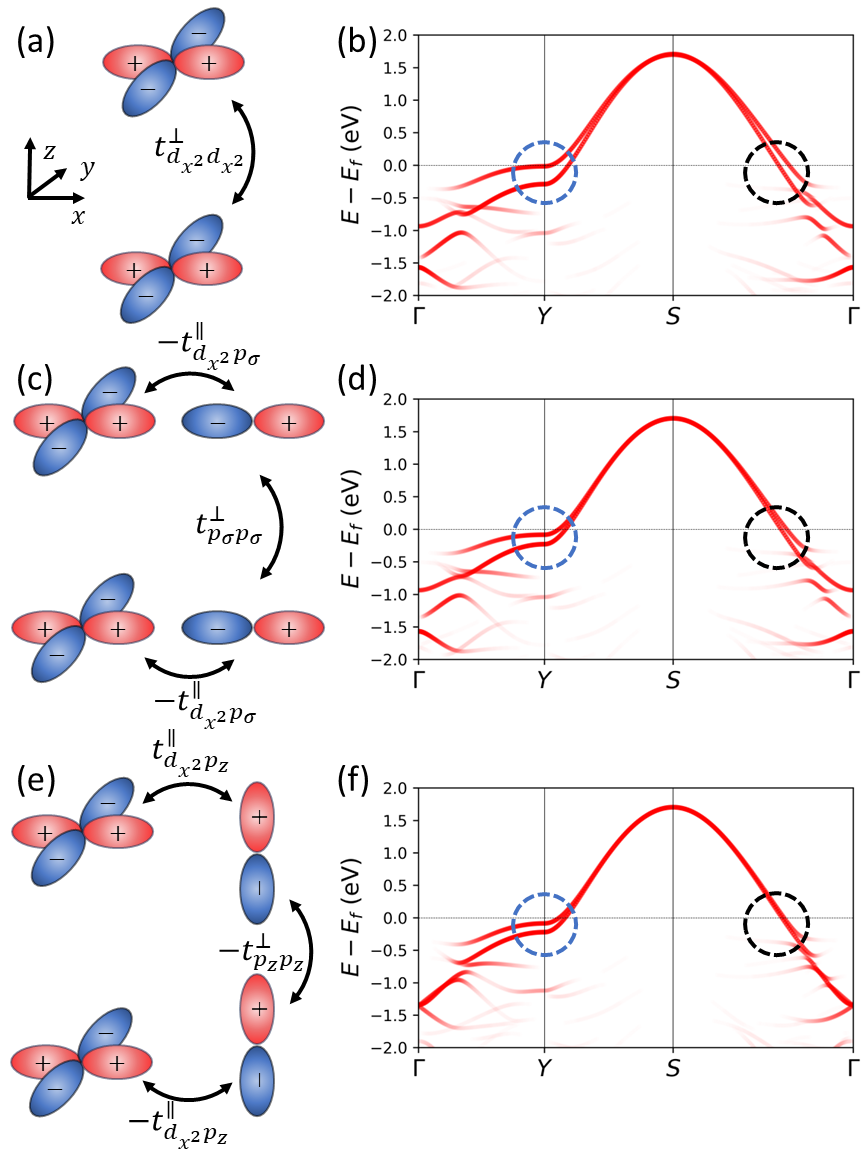}
\end{center}
\caption{
(a) Illustration of the direct interlayer hopping $t_{d_{x^2}d_{x^2}}^\perp$ between two $d_{x^2-y^2}$ orbitals from different layers.  (b) The unfolded Wannier band structure of YBCO7 where small hoppings weaker than 0.05 eV are truncated. The opacity of bands represents the projection weight to the planar $d_{x^2-y^2}$ orbitals.  The blue dashed circle on the left highlights the band splitting around the antinodal region originating from EIC, while the black dashed circle on the right highlights the nodal region. 
(c) Illustration of an effective interlayer coupling mechanism through the interlayer $p_{\sigma}$-$p_{\sigma}$ hopping.  (d) Band structure modified from (b) by setting $t_{p_{\sigma}p_{\sigma}}^\perp=0$.  (e) Illustration of an EIC mechanism through the interlayer $p_z$-$p_z$ hopping.  (f) Band structure modified from (d) by additionally setting $t_{p_zp_z}^\perp=0$.  
}
\label{fig:dppd}
\end{figure}
%%%%%%%%%%%%%%%%%%%%%%%%%%%%%%%%%%%%%%%%%%%%%%%%%%%%%%

The EIC between $d_{x^2}$ orbitals comes about due to many possible microscopic mechanisms each of which can contribute with different strengths.  In this section, we proceed systematically to identify the EIC mechanisms with the largest contributions. 

The simplest mechanism is the direct interlayer hopping between two $d_{x^2}$ orbitals illustrated in Fig.~\ref{fig:dppd}(a). We denote this hopping as $t_{d_{x^2}d_{x^2}}^\perp$ where the $d_{x^2}$ in the subscript represent two nearest neighbor Cu $d_{x^2}$ orbitals on two adjacent layers, and the superscript $\perp$ marks an interlayer hopping (as opposed to in-plane hopping marked by $\parallel$).  Any hopping strength can be read off from an off-diagonal matrix element of the Wannierized tight-binding Hamiltonian.  Among all the cuprates studied in this work, $|t_{d_{x^2}d_{x^2}}^\perp|$ is 0.01 to 0.02 eV: this is a very small contribution compared to $\Delta_{tot}$ which is usually one order of magnitude larger. For example, Fig.~\ref{fig:dppd}(b) shows the band structure of YBCO7 after truncating all hoppings smaller in magnitude than 0.05 eV so all hoppings $t_{d_{x^2}d_{x^2}}^\perp$ have been zeroed.  As highlighted by the blue and black dashed circles, the band splittings caused by EICs at the antinodal and nodal regions are still finite, about 0.1 to 0.2 eV, and much larger than the $t_{d_{x^2}d_{x^2}}^\perp$ hopping.  Hence this hopping channel can be safely ignored.

%%%%%%%%%%%%% TABLE %%%%%%%%%%%%%%%%%%%%%%%%%%%%%%%%%%%
\begin{table} 
\caption{
Wannierized hopping strengths (magnitudes) and their variability of all studied cuprates in this work: $t_{max}/t_{min}$ is the magnitude of the ratio of the largest to smallest hopping within the same class.
}
\begin{ruledtabular} 
\begin{tabular} { c c c}
Hopping & Strength (eV) & $t_{max}/t_{min}$\\
\hline
$t_{d_{x^2}p_z}^\parallel$ & 0.01$-$0.33 & 33 \\
$t_{d_{z^2}p_{\sigma}}^\perp$ & 0.06$-$0.14 & 2.3 \\
$t_{p_{\sigma}p_{\sigma}}^\perp$ & 0.10$-$0.22 & 2.2 \\
$t_{d_{x^2}d_{x^2}}^\perp$ & 0.01$-$0.02 & 2.0 \\
$t_{p_zp_z}^\perp$ & 0.31$-$0.59 & 1.9 \\
$t_{d_{z^2}p_{\sigma}}^\parallel$ & 0.30$-$0.42 & 1.4 \\
$t_{p_zp_z}^\parallel$ & 0.11$-$0.14 & 1.3 \\
$t_{d_{x^2}p_{\sigma}}^\parallel$ & 1.15$-$1.33 & 1.2 \\
$t_{p_{\sigma}p_{\sigma}}^\parallel$ & 0.59$-$0.65 & 1.1 \\
$t_{p_{\sigma}p_{\sigma}}^{\parallel,2^{nd}}$ & 0.09$-$0.11 & 1.2
\end{tabular} 
\end{ruledtabular} 
\label{tab:hops}
\end{table}

%%%%%%%%%%%%%%%%%%%%%%%%%%%%%%%%%%%%%%%%%%%%%%%%%%%%%%%

Apart from the direct hopping $t_{d_{x^2}d_{x^2}}^\perp$, $\Delta_{tot}$ can be mediated by various types of indirect interlayer hoppings.  Fig.~\ref{fig:dppd}(c) illustrates a mechanism where two layers are connected by the hopping between interlayer $p_{\sigma}$-$p_{\sigma}$ orbitals.  Such $t_{p_{\sigma}p_{\sigma}}^\perp$ hopping have magnitudes of 0.10 to 0.22 eV.  The $p_{\sigma}$ orbitals also serve as ``transport hubs'' for the EIC by their in-plane connections to the $d_{x^2}$ orbitals.  One example of such in-plane intralayer connections is the nearest-neighbor in-plane hopping between Cu and O atoms $t_{d_{x^2}p_{\sigma}}^\parallel$ as shown in Fig.~\ref{fig:dppd}(c).  Meanwhile, there are other in-plane intralayer connections such as a combination of $t_{d_{x^2}p_{\sigma}}^\parallel$ with nearest-neighbor $t_{p_{\sigma}p_{\sigma}}^\parallel$ or other longer-ranged hoppings discussed in Appendix~\ref{app:ppinpane}.  These different in-plane connections lead to multiple hopping pathways between two Cu $d_{x^2}$ orbitals sharing the same interlayer hopping path $t_{p_{\sigma}p_{\sigma}}^\perp$.   

In Table~\ref{tab:hops}, we list the scales of the most important hopping strengths in this work.  The second nearest neighbor in-plane hopping between $p_{\sigma}$ orbitals $t_{p_{\sigma}p_{\sigma}}^{\parallel,2^{nd}}$ is much smaller than the nearest neighbor hopping $t_{p_{\sigma}p_{\sigma}}^{\parallel}$, so we will ignore the contributions from long-range in-plane hoppings.  

Fig.~\ref{fig:dppd}(d) shows how the band structure of YBCO7 modified from Fig.~\ref{fig:dppd}(b) when we also zero the $t_{p_{\sigma}p_{\sigma}}^\perp$ hoppings.  The band splittings at both nodal and antinodal regions become much smaller compared to Fig.~\ref{fig:dppd}(b), indicating a large contribution from the $t_{p_{\sigma}p_{\sigma}}^\perp$ mechanism.  The change of band splittings going from Fig.~\ref{fig:dppd}(b) to (d) defines the contribution of this mechanism $\Delta_{p_{\sigma}p_{\sigma}\perp}$ which stems from all the hopping pathways sharing the same interlayer conduit $t_{p_{\sigma}p_{\sigma}}^\perp$.  In this way, we continue to numerically explore the contributions of other mechanisms by sequentially zeroing particular interlayer hoppings. 

Fig.~\ref{fig:dppd}(e) illustrates a $t_{p_zp_z}^\perp$ mechanism, where the $p_z$ orbitals are now the ones creating the interlayer connection.  The in-plane $t_{d_{x^2}p_z}^\parallel$ and $t_{p_zp_z}^\parallel$ hoppings create the final in-plane connection to the $d_{x^2}$ orbitals.  Note that for an ideal high-symmetry crystal without any distortions of the CuO$_2$ plane, the $t_{d_{x^2}p_z}^\parallel$ hopping will be zero by symmetry.  However, the oxygen atoms in the CuO$_2$ planes often move and create distortions along the $z$-axis, typically called ``buckling'' in previous literature \cite{piekarz1999electron, carbone2008direct, johnston2010systematic}.  Such distortions allow finite $t_{d_{x^2}p_z}^\parallel$ hoppings which turn out to play an important role in this mechanism.  By additionally zeroing out the $t_{p_zp_z}^\perp$ hoppings, the band structure in Fig.~\ref{fig:dppd}(d) changes to Fig.~\ref{fig:dppd}(f): the band splitting has further decreased and effectively vanishes at the nodal region (black dashed circle). 
%%%%%%%%%%% FIGURE %%%%%%%%%%%%%%%%%%%%%%%%%%%%%%%%%%%
\begin{figure}[t]
\begin{center}
\includegraphics[scale=0.32]{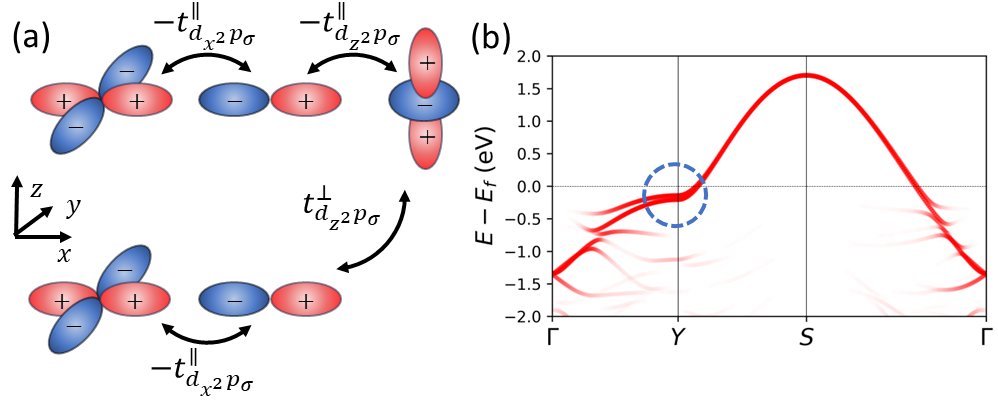}
\end{center}
\caption{
(a) Illustration of an EIC mechanism through the interlayer $d_{z^2}$-$p_{\sigma}$ hopping.  (b) The unfolded Wannier band structure modified from Fig.~\ref{fig:dppd}(f) by additionally setting $t_{d_{z^2} p_{\sigma}}^\perp=0$.  The opacity of bands represents the projection weight to the planar $d_{x^2-y^2}$ orbitals.  The blue dashed circle highlights how the band splitting changes around the antinodal region compared to Fig.~\ref{fig:dppd}(f). 
}
\label{fig:dpdpd}
\end{figure}
%%%%%%%%%%%%%%%%%%%%%%%%%%%%%%%%%%%%%%%%%%%%%%%%%%%%%%

Up to now, we have shown that the interlayer $p$-$p$ hoppings completely control the effective interlayer coupling at the nodal region.  For the antinodal region, they are obviously important but there must be some remaining important couplings since the antinodal splitting in Fig.~\ref{fig:dppd}(f) is still significant.  To complete the story, Fig.~\ref{fig:dpdpd}(a) illustrates a mechanism mediated by $t_{d_{z^2}p_{\sigma}}^\perp$ hopping.  By additionally zeroing these $t_{d_{z^2}p_{\sigma}}^\perp$, the band splitting at the antinodal region effectively becomes zero (compare Fig.~\ref{fig:dppd}(f) to Fig.~\ref{fig:dpdpd}(b)).  Of course, further minor interlayer or inter-bilayer mechanisms contribute to the remaining small band splittings.  As demonstrated in Appendix~\ref{app:mech}, these minor mechanisms are usually one order of magnitude smaller than the major mechanisms discussed above, and we will ignore them in the remainder of this work. 

\section{Hoppings and crystal structure}
\label{sec:hop_crystal}
%%%%%%%%%%% FIGURE %%%%%%%%%%%%%%%%%%%%%%%%%%%%%%%%%%%
\begin{figure*}[t]
\begin{center}
\includegraphics[scale=0.58]{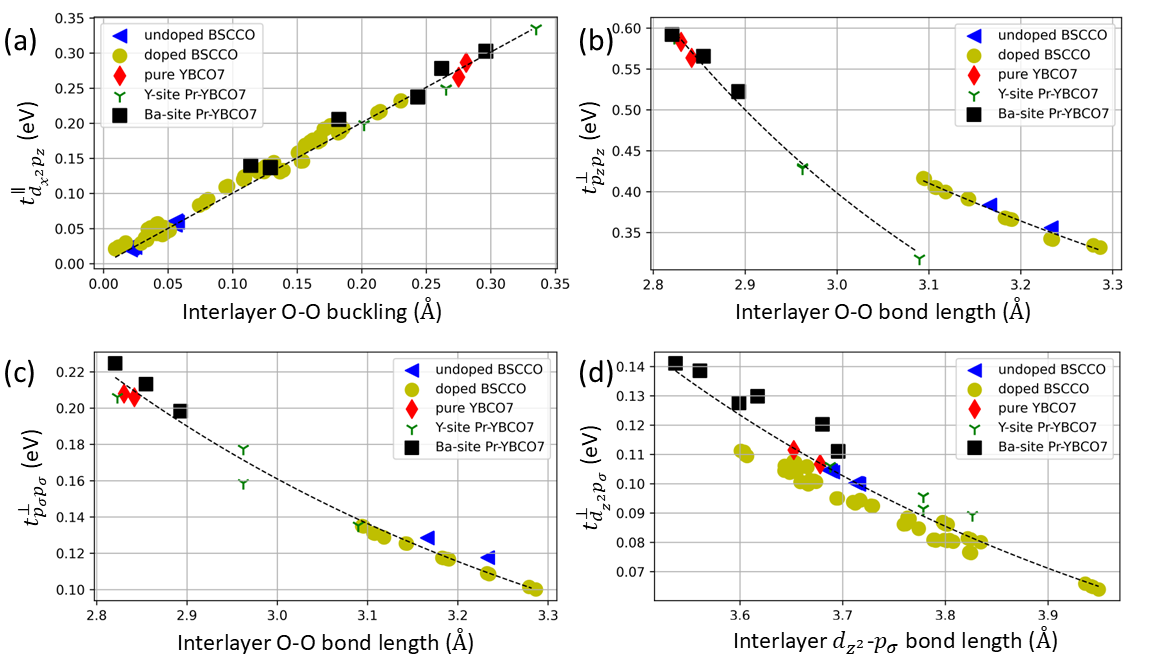}
\end{center}
\caption{
Relations between strongly-varying hoppings and structural properties.  Different data points are from symmetry-independent atoms in undoped BSCCO, 25\% hole-doped BSCCO, YBCO7, Y-site Pr-doped YBCO7, and Ba-site Pr-doped YBCO7. 
(a) In-plane hopping between $d_{x^2}$ and $p_z$ orbitals $t_{d_{x^2} p_z}^\parallel$ as illustrated in Fig.~\ref{fig:dppd}(e) versus the corresponding in-plane oxygen buckling strength.  The dashed line represents a linear relation fit.  (b) Inter-layer hopping between $p_z$ orbitals from different layers $t_{p_z p_z}^\perp$ as illustrated in Fig.~\ref{fig:dppd}(e) versus the interlayer oxygen-oxygen bond length.  (c) Inter-layer hopping between $p_{\sigma}$ orbitals from different layers $t_{p_{\sigma} p_{\sigma}}^\perp$ as illustrated in Fig.~\ref{fig:dppd}(c) versus the interlayer oxygen-oxygen bond length.  (d) Inter-layer hopping $t_{d_{z^2} p_{\sigma}}^\perp$ as illustrated in Fig.~\ref{fig:dpdpd}(a) versus the corresponding bond length. The dashed lines in (b-d) are given by exponential fits of the data points.  Detailed analytical forms of all analytical fits are tabulated in Appendix~\ref{app:formula}. % \red{We need horizontal vertical ticks marks and gridlines on these plots.}
}
\label{fig:hop_struct}
\end{figure*}
%%%%%%%%%%%%%%%%%%%%%%%%%%%%%%%%%%%%%%%%%%%%%%%%%%%%%%

Across the cuprates studied in this work, some of the hoppings participating in the EIC mechanisms depend strongly on the local crystal environment while others are less sensitive.  To make high-quality estimates of the interlayer hopping parameters based on the crystal structure, obviously one should focus first on the large hoppings that have strong structural dependence.  We sort the hoppings in Table~\ref{tab:hops} by their variability, and we will focus on several strongly varying and large hoppings in this section. 

As expected, the hopping strengths are closely related to the local structures around the atoms involved in the hoppings.  For example, an important component of the $p_z$-$p_z$ mechanism is the $t_{d_{x^2} p_z}^\parallel$ hopping, and Fig.~\ref{fig:hop_struct}(a) shows that $t_{d_{x^2} p_z}^\parallel$ is linearly dependent on the oxygen buckling magnitude, which is the out-of-plane displacement of the planar oxygen atom. 
Note that this linear relation is obeyed across the multiple materials studied: this transferability means that as long as the buckling strength is known either via DFT relaxation or experimental measurement, one can predict the hopping strength $t_{d_{x^2} p_z}^\parallel$ between any pair of adjacent CuO$_2$ layers.  Analytical formulae and fits for such relations are gathered and listed in Appendix~\ref{app:formula}.  

Another important participating hopping for the $p_z$-$p_z$ mechanism is the interlayer hopping  $t_{p_z p_z}^\perp$.  Note that the oxygen-oxgyen bond lengths between adjacent CuO$_2$ layers are usually quite large ($\sim 3$\AA) in these cuprates due to sandwiched atomic layers between two CuO$_2$ planes, e.g., a Ca layer in BSCCO and an Y layer in YBCO.  On top of direct oxygen-oxygen hopping processes between the CuO$_2$ layers, these sandwiched ions have ionized valence orbitals (4$s$ in Ca and $5s$ and $4d$ in Y) that can further bridge the interlayer hopping.  Since we have not explicitly included Wannier functions on these ions to arrive at a simple description, their effect is included implicitly in the Wannier Hamiltonian matrix elements that we are analyzing (i.e., they are projected out and renormalize the resulting hoppings).  As a result, $t_{p_z p_z}^\perp$ represents the effective interlayer hopping between $p_z$ orbitals that includes all such effects, and we plot it as a function of the interlayer O-O bond length in Fig.~\ref{fig:hop_struct}(b).  Due to the influence of the sandwich layers on this hopping path, we fit two different functions for the Ca and Y cases.    Next, Fig.~\ref{fig:hop_struct}(c) and (d) show $t_{p_{\sigma} p_{\sigma}}^\perp$ and $t_{d_{z^2} p_{\sigma}}^\perp$ as functions of the corresponding bond lengths.  The nature of sandwiched layer has a much smaller effect on these two types of interlayer hoppings than the one in Fig.~\ref{fig:hop_struct}(b).  In all three cases, we fit the data with exponentially decaying curves (see Appendix~\ref{app:formula} for details).

The fitted functions shown in Figs.~\ref{fig:hop_struct}(a-d) allow us to easily estimate the hoppings using only simple structural parameters.  Of course, as visible in the figures, there are finite errors involved with these estimates.  (We tabulate the standard deviations of the data away from the fits in each case to estimate errors in predicted EIC further below.)  In principle, one could include more structural properties in the estimations to reduce the errors.  For example, we find that including the bond angles and orbital orientations using the Slater and Koster method \cite{slater1954simplified} can slightly reduce errors by about 10\%.  As the improvements seem very modest, for simplicity, we only focus on the dominant structural properties in this work to avoid complicated analytical formulae and use only fits that depend on a single structural variable (as shown in Figs.~\ref{fig:hop_struct}(a-d)).

In addition the hoppings discussed above, there are several less variable hoppings ($t_{max}/t_{min}\leq 1.4$) participating in the EIC mechanisms.  Not surprisingly, the corresponding bond lengths also display low variability.  This allows us to use constants to estimate some of these hopping strengths with less than $10\%$ resulting error.  Detailed discussions of these hoppings are found in Appendix~\ref{app:lessvary}, and analytical formulae are tabulated in Appendix~\ref{app:formula}.  More accurate estimations of these hoppings would require a more complex material-specific understanding beyond simple structural information.  Here we opt for simplicity and instead keep track of accumulated errors throughout our estimations.  

\section{Estimating effective interlayer coupling}
\label{sec:estimate_EIC}

In the previous section, we described simple fits that allow rapid estimation of the key tight-binding hopping parameters controlling the EIC.  One could use these estimated hoppings to construct a tight-binding Hamiltonian and diagonalize it numerically to find the interlayer splitting.  However, for practical analysis of experimental data, it is preferable to have simple analytical forms that are easy to plot and (re)fit to measurements.  

Theoretically, the straightforward path to do this is to treat the hoppings using perturbation theory starting from the on-site only part of our Wannierized $d$-$p$ model.
When all hoppings are treated  perturbatively to the lowest order, the EICs from different mechanisms add 
\begin{equation}
\Delta_{tot} = \sum_M \Delta_M
\label{equ:delta_tot}
\end{equation}  
where $\Delta_{tot}$ is the total band splitting caused by EICs, and $\Delta_M$ is the interlayer band splitting from a microscopic mechanism $M$.  By different mechanisms, we mean physically distinct hopping pathways between two Cu $d_{x^2}$ orbitals as exemplified by Figs.~\ref{fig:dppd}(a,c,e) or Fig.~\ref{fig:dpdpd}(a).  

%%%%%%%%%%% FIGURE %%%%%%%%%%%%%%%%%%%%%%%%%%%%%%%%%%%
\begin{figure}[t]
\begin{center}
\includegraphics[scale=0.41]{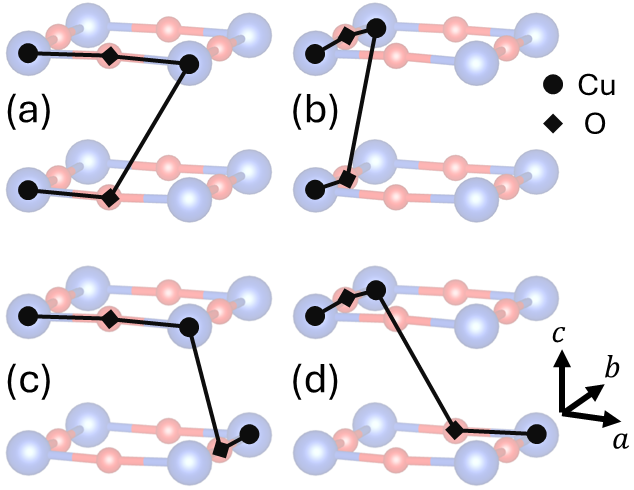}
\end{center}
\caption{
Multiple hopping pathways for the mechanism mediated by interlayer $d_{z^2}$-$p_{\sigma}$ hopping as illustrated in Fig.~\ref{fig:dpdpd}(a).  These four hopping pathways are classified by the directions of the upper-layer $d_{x^2}$-$p_{\sigma}$-$d_{z^2}$ hopping and the lower-layer $p_{\sigma}$-$d_{x^2}$ hopping, namely (a) $\pm x$ and $\pm x$; (b) $\pm y$ and $\pm y$; (c) $\pm x$ and $\pm y$; (d) $\pm y$ and $\pm x$ directions.  The background light blue and red atoms and bonds show the bilayer structure of BSCCO.  The black circles and diamonds represent a path involving Cu and O atoms.    
}
\label{fig:pathway}
\end{figure}
%%%%%%%%%%%%%%%%%%%%%%%%%%%%%%%%%%%%%%%%%%%%%%%%%%%%%%

%%%%%%%%%%% FIGURE %%%%%%%%%%%%%%%%%%%%%%%%%%%%%%%%%%%
\begin{figure*}[t]
\begin{center}
\includegraphics[scale=0.41]{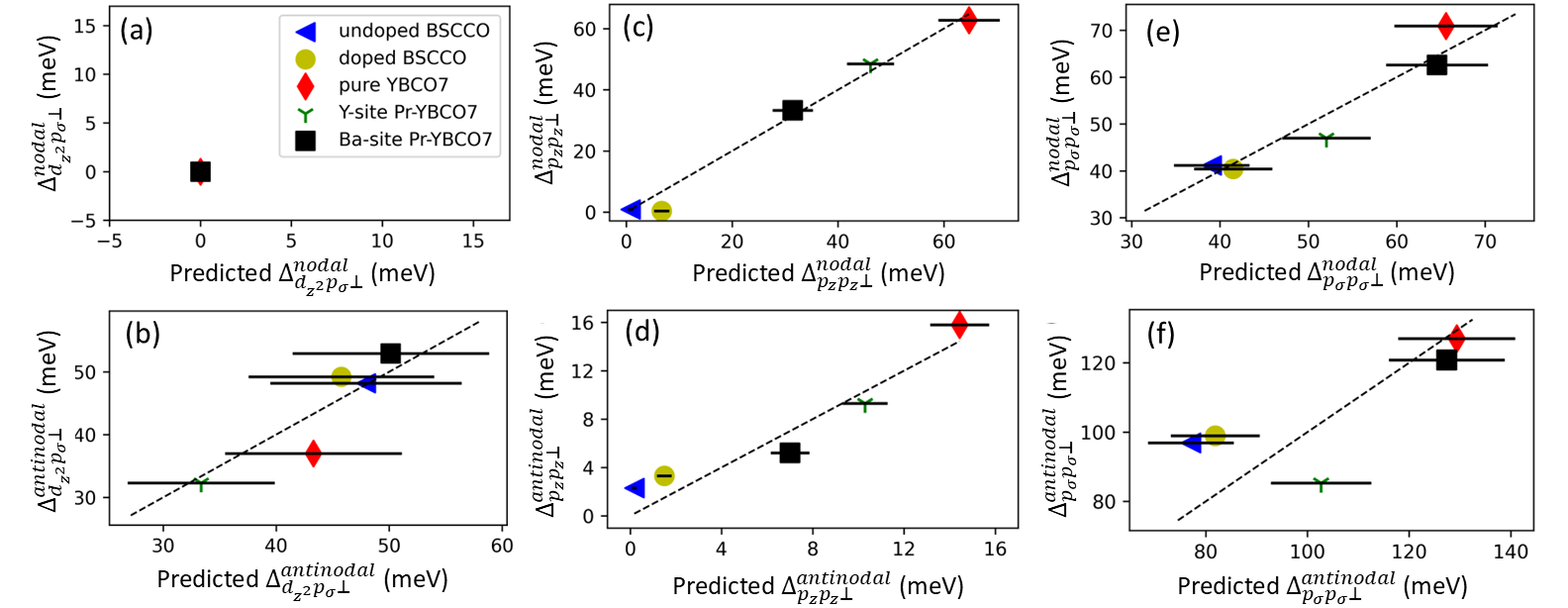}
\end{center}
\caption{
Perturbation estimations of EICs from structural properties.
% (\red{once you have finished writing the formulae, you should refer to one of them here so they know what you are referring to}).  
The vertical axis shows the EIC computed from the Wannier model.  The horizontal axis represents EIC predicted by Eq.~\eqref{equ:dm_general}.  The nodal point refers to $(0.42\pi/a, 0.42\pi/a, 0)$ in the first BZ, while the antinodal point refers to $(0.18\pi/a, \pi/a, 0)$ in the first BZ.
(a) Computed contribution of the $d_{z^2}p_{\sigma}\perp$ mechanism at the nodal point $\Delta_{d_{z^2}p_{\sigma}\perp}^{nodal}$ versus the formula-predicted result from the structural properties.  Blue triangle, yellow circle, red diamond, green ``Y'' symbol, and black square represent undoped BSCCO, doped BSCCO, pure YBCO7, Y-site Pr-YBCO7, and Ba-site Pr-YBCO7, respectively.  Specifically, $\Delta_{d_{z^2}p_{\sigma}\perp}^{nodal}$ is always zero, so all data points overlap with each other.  (b) Contributions of the $d_{z^2}p_{\sigma}\perp$ mechanism at the antinodal point.  (c-d) Comparisons of the $p_zp_z\perp$ mechanisms at the nodal and antinodal points.    (e-f) Comparisons of the $p_{\sigma}p_{\sigma}\perp$ mechanisms at the nodal and antinodal points.   
}
\label{fig:delta_predict_hop}
\end{figure*}
%%%%%%%%%%%%%%%%%%%%%%%%%%%%%%%%%%%%%%%%%%%%%%%%%%%%%%

When we consider the mechanism shown in Fig.~\ref{fig:dpdpd}(a), we note that there are multiple possible pathways for the in-plane hopping within this mechanism (i.e., along $\pm x$, $\pm y$, or other directions for longer-ranged hopping). Fig.~\ref{fig:pathway} shows four different pathways of this same mechanism mediated by $t_{d_{z^2}p_{\sigma}}^{\perp}$, where both the upper-layer $d_{x^2}$-$p_{\sigma}$-$d_{z^2}$ hopping and the lower-layer $p_{\sigma}$-$d_{x^2}$ hopping may go along $\pm x$ or $\pm y$ directions.  The EICs of different pathways are also additive in perturbation theory, so we have 
\begin{equation}
\Delta_{M} = \sum_{p\in M} \Delta_p
\label{equ:delta_m_sum}
\end{equation}
where $p$ represents different hopping pathways in a given mechanism $M$.  

When examining a particular pathway $p$, we see that, within lowest order perturbation theory, the general form for $\Delta_p$ will be of a numerator composed of products of hoppings and a denominator made of a product of energy differences,
\begin{equation}
\Delta_p = \frac{\prod_{s} t_s}{\prod_{i} E_i}\,,
\end{equation}
where the path $p$ is written as a sequence of elementary steps $s$ with hopping $t_s$ (e.g., the specific path shown in Fig.~\ref{fig:pathway}(a)) and the $E_i$ are the corresponding intermediate state energy denominators. 

In principle, one could attempt to compute or estimate the individual energy denominators which would require materials-specific knowledge of the local onsite energies and their dependence on the local environment (e.g., local structural distortions).  However, to keep our scheme simple and general, we will make the approximation that the product of energy denominators for all the pathways of a given mechanism is a universal fixed number. This simplifies $\Delta_p$ to
\begin{equation}
\Delta_p = c_M\prod_{s} t_s\,,
\label{equ:prod_ts}
\end{equation}
where $c_M$ is a mechanism-dependent constant for a pathway $p\in M$.  Now, we denote each mechanism by the key hoppings involved.  For example, $t_{d_{x^2}p_{\sigma}}^\parallel t_{p_{\sigma}d_{z^2}}^\parallel t_{d_{z^2}p_{\sigma}}^\perp t_{p_{\sigma}d_{x^2}}^\parallel$ represents the mechanism mediated by $t_{d_{z^2}p_{\sigma}}^\perp$ interlayer hopping as illustrated in Fig.~\ref{fig:dpdpd}(a), and we will use this particular example to describe our workflow. The numerical values of the hopping strengths can be estimated from structural properties as discussed in the previous section, but we must also account for the different phases of each of the hopping steps in a given pathway.  For example, the first step of the pathway Fig.~\ref{fig:dpdpd}(a) involves tunneling from Cu $dx^2$ to the neighboring O $p_\sigma$ along the $+x$ direction: the $t_s$  for this step is hopping including its phase is $t_{d_{x^2}p_{\sigma}}^\parallel e^{ik_xa/2}$.  For this same elementary step going along $-x$, one would have instead $-t_{d_{x^2}p_{\sigma}}^\parallel e^{-ik_xa/2}$. 

Given the above assumptions and observations, we can now write $\Delta_M$ as 
\begin{equation}
\Delta_M = c_M D_M(k)\prod_{h\in M} t_h\,,
\label{equ:dm_general}
\end{equation}
where the string of hopping denoted by $h$ is the same as the name of the mechanism: e.g., for the mechanism exemplified Fig.~\ref{fig:dpdpd} the string of hoppings  $t_{d_{x^2}p_{\sigma}}^\parallel t_{p_{\sigma}d_{z^2}}^\parallel t_{d_{z^2}p_{\sigma}}^\perp t_{p_{\sigma}d_{x^2}}^\parallel$.  The overall factor $D_M(k)$ is the sum of the product of phases over the various pathways for that mechanism. 

In principle, when building up the product of shopping in Eq.~(\ref{equ:dm_general}), we should use our structure-based fits (i.e., the dashed curves in Fig.~(\ref{fig:hop_struct}) to compute each specific hopping based on its local environment which leads to a complex and purely numerical evaluation.  Since one of our primary goals is to have simple and analytic (or close to analytic) expressions for $\Delta_M$, we will make a simplifying assumption: we use a single numerical value for a particular hopping step (e.g., a single $t_{d_{x^2}p_{\sigma}}^\parallel$ value for all $d_{x^2}$ to $p_\sigma$ shopping), and the value we use is the average one over all the instances occurring in the unit cell.  This averaging over local structural variations gives us the desired mean value $\bar t_h$ but also the associated standard deviation $\sigma_h$.  Separately, the fitted relations shown in Fig.~(\ref{fig:hop_struct}) are not perfect, and this standard deviation of the fits away from the data for a given hopping is denoted as $\sigma_h'$.  For purposes of error propagation below, we will assume the two fluctuations add in quadrature so we expect the actual hopping to deviate from the mean by
\begin{equation}
t_h = \bar t_h \pm \delta t_h\, \ \ , \ \ \delta t_h=\sqrt{\sigma_h^2+\sigma_h^{\prime2}}\,.
\label{equ:error_th}
\end{equation}
Appendix~\ref{app:formula} tabulates $\sigma_h'$ for each hopping type.

By substituting Eq.~\eqref{equ:error_th} into equation Eq.~(\ref{equ:dm_general}), the errors of hoppings propagate to EICs as 
\begin{equation}
\delta\Delta_M = \bar\Delta_M \cdot\sqrt{\sum_{h\in M}(\delta t_h/\bar t_h)^2}\,,
\label{equ:error_dm}
\end{equation}
where the mean value $\bar\Delta_M$ is defined by substituting $\bar t_h$ into the hopping strengths in Eq.~\eqref{equ:dm_general}. 

We illustrate the computation of $D_M(k)$ for our chosen example. 
Considering Fig.~\ref{fig:dpdpd}(a), as noted above one can go along $\pm x$ for the first hopping involved, and the two possibilities for this particular subcase sum to 
$e^{ik_xa/2}-e^{-ik_xa/2}=2i \sin(k_xa/2)$.  Repeating this process over all possible pathways equivalent to Fig.~\ref{fig:dpdpd}(a) gives a contribution to $\Delta_M$ of 
\begin{multline*}
c_M\left(2i\sin\left(k_xa/2\right) t_{d_{x^2}p_{\sigma}}^\parallel\right)
\left(2i\sin\left(k_xa/2\right) t_{p_{\sigma}d_{z^2}}^\parallel\right)\\
\left(-2i\sin\left(k_xa/2\right) t_{d_{z^2}p_{\sigma}}^\perp\right)
\left(-2i\sin\left(k_xa/2\right) t_{d_{x^2}p_{\sigma}}^\parallel\right)\\
=16c_M\sin^4\left(k_xa/2\right) t_{d_{x^2}p_{\sigma}}^{\parallel 2}  t_{p_{\sigma}d_{z^2}}^\parallel  t_{d_{z^2}p_{\sigma}}^\perp
\end{multline*}
so the contribution to $D_M(k)$ is $16\sin^4(k_xa/2)$.    Repeating this process for each of the four different classes of pathways shown in Fig.~\ref{fig:pathway} gives us contributions to $D_M(k)$ of $16\sin^4(k_ya/2)$ for (b) and $-16\sin^2(k_xa/2)\sin^2(k_ya/2)$ for (c) and (d).  The $k$-dependencies of other cases and mechanisms are detailed in Appendix~\ref{app:formula}.  The final summed up result for $D_M(k)$ for this mechanism is
$D_M(k) = \left[\sin^2\left(k_xa/2\right)- \sin^2\left(k_ya/2\right)\right]^2$
giving the following form for the EIC contribution for this mechanism:
\begin{multline}
\Delta_M(k) = c_M\left[\sin^2\left(k_xa/2\right)- \sin^2\left(k_ya/2\right)\right]^2\\ t_{d_{x^2}p_{\sigma}}^{\parallel 2}  t_{p_{\sigma}d_{z^2}}^\parallel  t_{d_{z^2}p_{\sigma}}^\perp
\label{equ:delta_m}
\end{multline}
(we absorbed the overall coefficient of 16 into $c_M$).  Similarly, we have derived the form of $D_M(k)$ for other mechanisms (see Appendix~\ref{app:formula}).  The numerical values for $c_M$ for five cuprate materials are fit at their nodal and antinodal points, which provide ten data points to fit the coefficient for each mechanism.  The fits work reasonably well as shown in Fig.~\ref{fig:delta_predict_hop}.  The numerical values of $c_M$ are listed in Appendix~\ref{app:formula}.

We are now ready to test our final simple analytical forms for $\Delta_M(k)$ against 
the actual DFT-based numerical values of $\Delta_M(k)$ from the sequential zeroing process of Section \ref{sec:zeroing}. 
Fig.~\ref{fig:delta_predict_hop} compares our formula-predicted $\Delta_M(k)$ (horizontal axis) with the corresponding value given by direct calculations (vertical axis).  Figs.~\ref{fig:delta_predict_hop}(a,b) are for the $d_{z^2}p_{\sigma}\perp$ mechanisms at the nodal and antinodal regions, respectively.  Each data point comes with an error bar due to the approximations in the formula-based prediction, which is computed by Eq.~\eqref{equ:error_dm}.  
% \red{I don't understand your description of the error bars and how you got them: you tell me why you have error bars but not how you found the error bars.}  
The dashed lines are references with unit slope through the origin.  (We note that the EIC of the $d_{z^2}p_{\sigma}\perp$ mechanism is always zeros at the nodal region by symmetry.)  Similarly, Figs.~\ref{fig:delta_predict_hop}(c-f) show the predicted and the DFT-based $\Delta_M$ for the $p_{\sigma}p_{\sigma}\perp$ and $p_zp_z\perp$ mechanisms.  The general observation is that the predictions do well for the overall trends as well as the slope, and the worst-case errors are 10-20 meV.

% \section{Example usage: YBCO7}
\section{Predicted EIC throughout $k$-space}
\label{sec:example}

%%%%%%%%%%% FIGURE %%%%%%%%%%%%%%%%%%%%%%%%%%%%%%%%%%%
\begin{figure}[t]
\begin{center}
\includegraphics[scale=0.4]{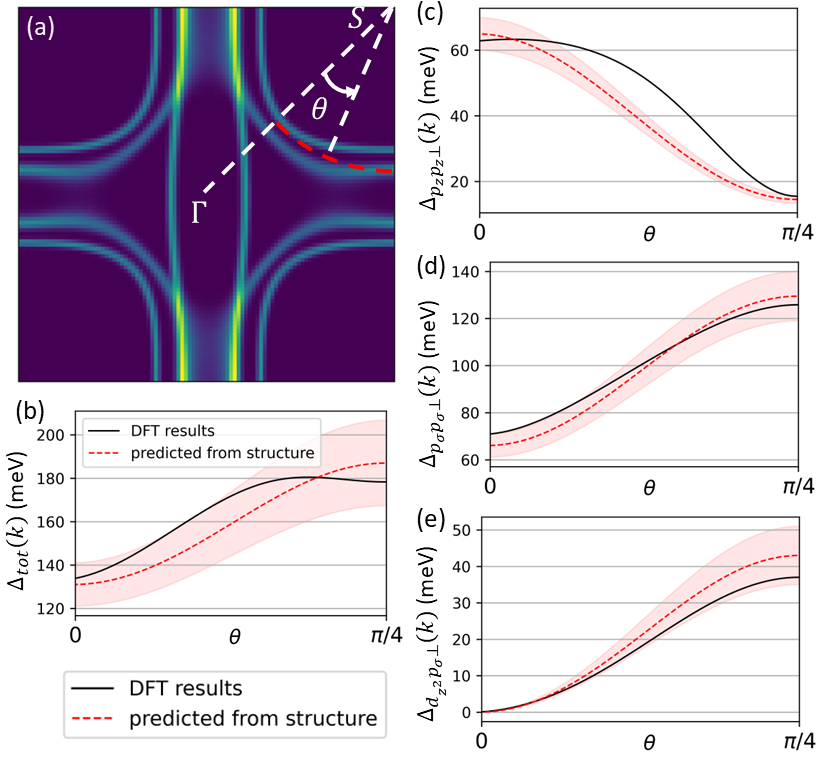}
\end{center}
\caption{
Benchmark estimation of major EIC mechanisms in YBCO7.  (a) The DFT Fermi surface, where the red dashed curve illustrates the $k$-path to be studied.  The path is along a circle with a radius of $0.41\pi/a$ centered at the $S$ point ($\pi/a$, $\pi/a$).   
The angle $\theta$ increases from 0 (nodal region) to $\pi/4$ (antinodal region).  (b) The total interlayer splitting as a function of $k$, which consists of three dominant microscopic mechanisms, namely the (c) $p_zp_z\perp$, (d) $p_{\sigma}p_{\sigma}\perp$, and (e) $d_{z^2}p_{\sigma}\perp$ mechanisms.  Red dashed curves show the predicted EICs based on the crystal structure, while the black solid curves show the actual results from DFT calculations.  The red shadows show the predicted error bars computed from Eq.~\eqref{equ:error_dm}.  
%\blue{The $k$-dependencies of the error bars follows $D_M(k)$. } 
% \red{I don't understand what this means: the error bars definitely follow $D_M(k)$ but I have no idea what it means to have these errors bars, where they came from, and how their k-dependence was determined.} 
}
\label{fig:hop_kdep}
\end{figure}
%%%%%%%%%%%%%%%%%%%%%%%%%%%%%%%%%%%%%%%%%%%%%%%%%%%%%
In this section, we provide an example of predicting the EIC in YBCO7 using our methodology: the prediction is not limited to nodal and antinodal regions but is performed for any $k$-point in the Brillouin zone (BZ).  We will benchmark the $k$-dependence of the EIC between nodal and antinodal regions following the main arcs of the Fermi surface shown in Fig.~\ref{fig:hop_kdep}(a). 

The crystal structure information is obtained from the relaxed DFT structure.  Based on the structural properties, we estimate the hoppings and the $\Delta_M$ (EICs) for the major microscopic mechanisms discussed above.  Detailed formulae for the different mechanisms are listed in Appendix~\ref{app:formula}.  The contributions from all mechanisms are added up to obtain $\Delta_{tot}=\sum_M \Delta_M$.  Fig.~\ref{fig:hop_kdep}(b) shows $\Delta_{tot}$ predicted from structural properties and compares it with the actual DFT results.  The deviations of the predictions from the DFT results are usually under 10 meV regardless of the $k$-point, smaller than the error bars.  

More precise predictions than those provided by our scheme would require more detailed knowledge of the materials systems.  For example, the orthorhombic symmetry breaking introduced by the oxygen dopants in YBCO7 is not taken into consideration in our current estimation method.  A precise description of the orthorhombic effect requires the onsite energy difference of planar oxygens along the $a$ and $b$ axis which is not easy to access from experiments.  Our DFT results that include orthorhombicity show that $\Delta_{tot}$ at one antinodal region ($\theta=-\pi/4$) is around 190 meV which is about 15 meV larger than for the antinodal region at $\theta=\pi/4$. By contrast, our simplified averaging approach for the hoppings yields the same value of $\Delta_{tot}=183$ meV for both cases (which is the average of the two more accurate values).  We find this level of error to be acceptable given the simplicity of the approach.

In addition, the nodal region ($\theta=0$) displays a smaller $\Delta_{tot}$ than the antinodal region ($\theta=\pi/4$), consistent with experimental findings \cite{zabolotnyy2007momentum}.  This is a common phenomenon for bilayer cuprates, such as BSCCO \cite{he2021superconducting}.  The most popular empirical fitting formula (Eq. \eqref{equ:old}) used in prior literature can also capture this behavior.  However, the EIC is not always larger at the antinodal region for all microscopic mechanisms.  Fig.~\ref{fig:hop_kdep}(c-e) show the individual contributions of the major mechanisms, i.e., mediated by interlayer $p_zp_z$, $p_{\sigma}p_{\sigma}$, and $d_{z^2}p_{\sigma}$ hoppings.  While $\Delta_{p_{\sigma}p_{\sigma}\perp}$ and $\Delta_{d_{z^2}p_{\sigma}\perp}$ show the same $k$-dependent behavior as $\Delta_{tot}$ in $k$-space, $\Delta_{p_zp_z\perp}$ is smaller at the antinodal region compared to the nodal region.  Note that our predictions correctly capture the separate $k$-dispersion for these major mechanisms. 

\section{Conclusions}
Using DFT studies of a few cuprates, we could reproduce their experimental ARPES spectra.  With the help of the Wannierized tight-binding model extracted from the DFT calculations, we revealed various microscopic mechanisms of EICs between planar Cu $d_{x^2}$ orbitals.  We computed quantitatively the contributions of these mechanisms and found that the major mechanisms are mediated by interlayer $p_zp_z$, $p_{\sigma}p_{\sigma}$, and $d_{z^2}p_{\sigma}$ hoppings.  In addition, we found that the hopping strengths are directly related to structural properties.  We have shown that such intimate relations are mostly material-independent and can even be described by simple and generalizable analytical formulae.  Based on these relations, we have derived a perturbative analytical estimator for the EIC based on structural properties alone.  This structure-based estimation was benchmarked for various $k$-points in the first BZ and showed acceptable errors compared to DFT results.  

In contrast to prior empirical fittings of the EIC in cuprates, our work provides a more robust microscopic understanding that can be generalized across different materials.  The structure-based estimation formulae can help predict and understand the EICs using only structural information.  The structural properties, either from DFT calculations or experiments, are usually more accessible than the detailed electronic spectrum, especially for materials such as YBCO where clean-surface cleavage is non-trivial and thus high-quality electron spectroscopy is difficult to perform.  In addition, our work provides analytical $k$-dependences for the different microscopic mechanisms which can improve the accuracy of tight-binding models fit to ARPES experiments.  An interesting future work direction involves generalizing these microscopic mechanisms and corresponding formulae to understand effective interlayer couplings between planar orbitals in other layered metal oxides beyond cuprates.

\section{Acknowledgements}
We thank Yu He, Jinming Yang, and Siqi Liu for their helpful discussions. This work was supported by grant NSF DMR 2237469, NSF ACCESS supercomputing resources via allocation TG-MCA08X007, and computing resources from Yale Center for Research Computing.

\appendix
\section{Computational details}
\label{app:DFT}
All crystal and electronic structure calculations in this study were conducted using the 5.4.4 version of the VASP plane-wave DFT software \cite{kresse1996efficiency, kresse1996efficient} with the PBE+U functional. To strike a balance between computational efficiency and accuracy, we aimed to converge the energy differences between various crystal or spin structures to within 1 meV per Cu atom. Based on previous studies \cite{jin2023afm, zhang2020competing}, we adopted a relatively high cutoff energy of 500 eV for the plane-wave basis set. All VASP calculations were performed with an energy tolerance of EDIFF $= 10^{-5}$ eV and a force tolerance of EDIFFG $= -1 \times 10^{-2}$ eV/Å. For SCF calculations, a Gaussian smearing of 0.2 eV was applied which is the largest smearing value that keeps the (artificial) smearing entropy contribution $E-F<1$ meV per Cu.

%%%%%%%%%%% FIGURE %%%%%%%%%%%%%%%%%%%%%%%%%%%%%%%%%%%
\begin{figure*}[t]
\begin{center}
\includegraphics[scale=0.5]{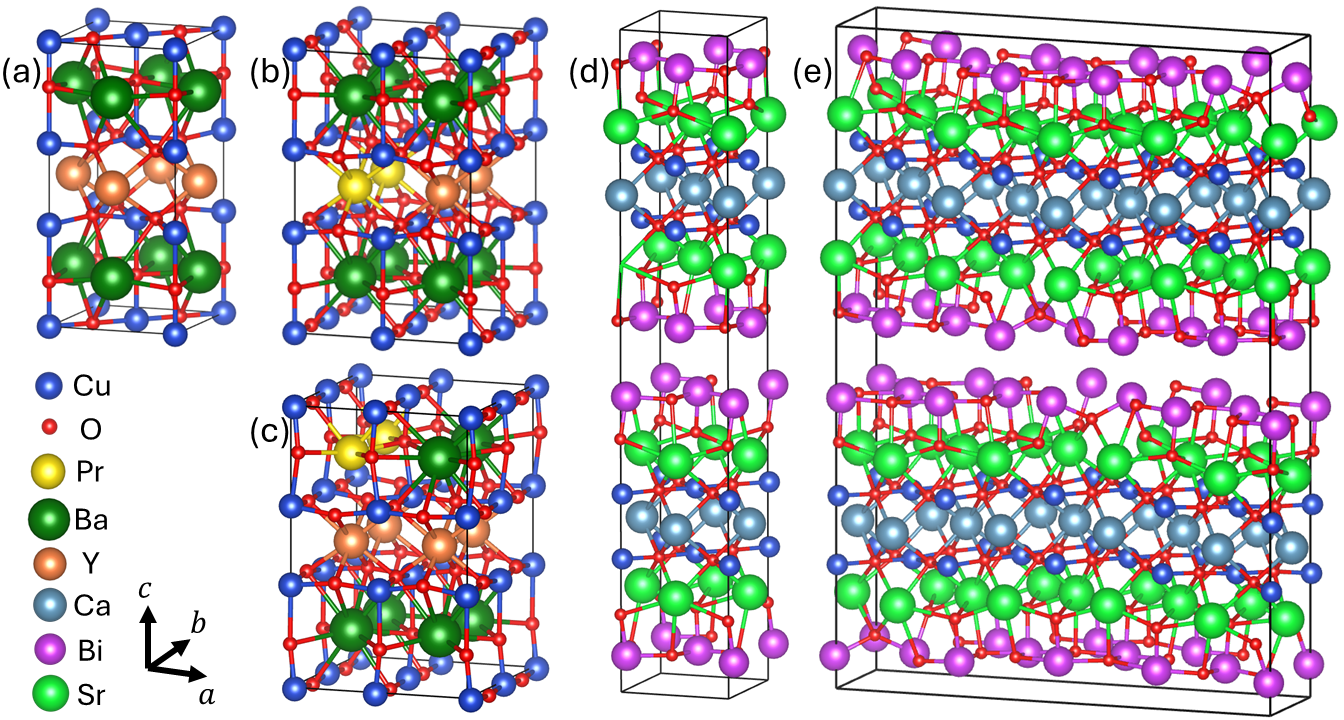}
\end{center}
\caption{
DFT relaxed lowest-energy crystal structures of (a) YBa$_2$Cu$_3$O$_7$, (b) Pr$_{0.5}$Y$_{0.5}$Ba$_2$Cu$_3$O$_7$, (c) Pr$_{0.5}$YBa$_{1.5}$Cu$_3$O$_7$, (d) Bi$_2$Sr$_2$CaCu$_2$O$_8$, and (e) Bi$_2$Sr$_2$CaCu$_2$O$_{8.25}$. 
}
\label{fig:crystal}
\end{figure*}
%%%%%%%%%%%%%%%%%%%%%%%%%%%%%%%%%%%%%%%%%%%%%%%%%%%%%

%%%%%%%%%%%%% TABLE %%%%%%%%%%%%%%%%%%%%%%%%%%%%%%%%%%%
\begin{table*} 
\caption{
The chemical formula, supercell size, lattice constants, and $k$-mesh of each cuprate system in Fig.~\ref{fig:crystal}.
}
\begin{ruledtabular} 
\begin{tabular} { c c c c c c c}
Label & Chemical formula & Supercell size & $a$ (\AA) & $b$ (\AA) & $c$ (\AA) & $k$-mesh \\
\hline
(a) & YBa$_2$Cu$_3$O$_7$ & $\sqrt{2}\times\sqrt{2}\times1$ & 5.49 & 5.49 & 11.80 & $8\times8\times4$\\
(b) & Pr$_{0.5}$Y$_{0.5}$Ba$_2$Cu$_3$O$_7$ & $2\times2\times1$ & 7.69 & 7.88 & 11.77 & $6\times6\times4$\\
(c) & Pr$_{0.5}$YBa$_{1.5}$Cu$_3$O$_7$ & $2\times2\times1$ & 7.63 & 7.83 & 11.72 & $6\times6\times4$\\
(d) & Bi$_2$Sr$_2$CaCu$_2$O$_8$ & $\sqrt{2}\times\sqrt{2}\times1$ & 5.43 & 5.50 & 31.98 & $8\times8\times1$\\
(e) & Bi$_2$Sr$_2$CaCu$_2$O$_{8.25}$ & $4\sqrt{2}\times\sqrt{2}\times1$ & 21.79 & 5.42 & 31.70 & $3\times8\times1$\\

\end{tabular} 
\end{ruledtabular} 
\label{tab:kpoints}
\end{table*}

%%%%%%%%%%%%%%%%%%%%%%%%%%%%%%%%%%%%%%%%%%%%%%%%%%%%%%%

The minimum required $k$-mesh density varies depending on the system size.  Fig.~\ref{fig:crystal} shows the unit cells of the five different cuprate systems examined in this study.  In contrast to the primitive cell, which contains one Cu atom per layer, we utilized larger supercells to allow for energy-lowering structural distortions.  Previous studies have demonstrated that incorporating these realistic distortions is crucial for accurately describing the key structural, electronic, and magnetic properties of the cuprates' normal state \cite{jin2023afm}.  The sizes of the supercells used in this work are detailed in Table \ref{tab:kpoints}.  Based on earlier works \cite{jin2023afm, zhang2020competing}, we selected relatively dense $k$-meshes for each cuprate system, as listed in Table \ref{tab:kpoints}.

For both undoped and doped BSCCO systems, we adopted and then relaxed structures from our previous work \cite{jin2023afm} which are consistent with high-resolution scanning transmission electron microscopy (STEM) measurements \cite{song2019visualization}.  For YBCO7 systems, we followed the structures from prior studies \cite{zhang2020competing} and performed additional structural relaxations within our computational framework.  The relaxed lattice constants of both BSCCO and YBCO in Table \ref{tab:kpoints} show small errors $<1\%$ compared to corresponding experimental measurements \cite{poole1989structural}.  In the case of Pr doping, we applied $U_{Pr} = 8$ eV to the $f$-orbitals and substituted Pr into either the Y or Ba sites. In both substitution cases, two Pr dopants were placed in the supercell, resulting in multiple possible dopant configurations.  We relaxed the structures for all potential dopant positions in the G-AFM states, and the lowest-energy configurations are shown in Fig.~\ref{fig:crystal}(b) and (c).  In all cases, we find robust Pr$^{3+}$ cations with energetically narrow occupied and unoccupied $f$-bands that lie at least 3 eV away from the Fermi energy. 

For Y-site substitution, Pr dopants tend to align along the $b$-axis (lowest energy), with alternative meta-stable alignments along the $a$-axis (1 meV higher) or the diagonal (110) direction (85 meV higher).  Due to the orthorhombic distortion introduced by the CuO chain along the $b$-axis, Pr dopants show a slight energetic preference for alignment along the $b$-axis compared to the $a$-axis.  However, this energy difference is small because of the relatively large distance between the CuO chain layer and the Y layer.  In contrast, for Ba-site substitution, which is much closer to the CuO chain layer, the energy difference is more pronounced: alignment along the $a$-axis is approximately 0.65 eV higher in energy than alignment along the $b$-axis.  Given that the unit cell contains two Ba layers, there are seven possible dopant configurations for two Pr atoms.  The lowest-energy structure still favors $b$-axis alignment, as shown in Fig.~\ref{fig:crystal}(c).  The second lowest-energy structure, with two Pr atoms aligned along the $c$-axis, is 0.12 eV higher in energy than the lowest-energy configuration and is not considered.

%%%%%%%%%%% FIGURE %%%%%%%%%%%%%%%%%%%%%%%%%%%%%%%%%%%
\begin{figure*}[t]
\begin{center}
\includegraphics[scale=0.35]{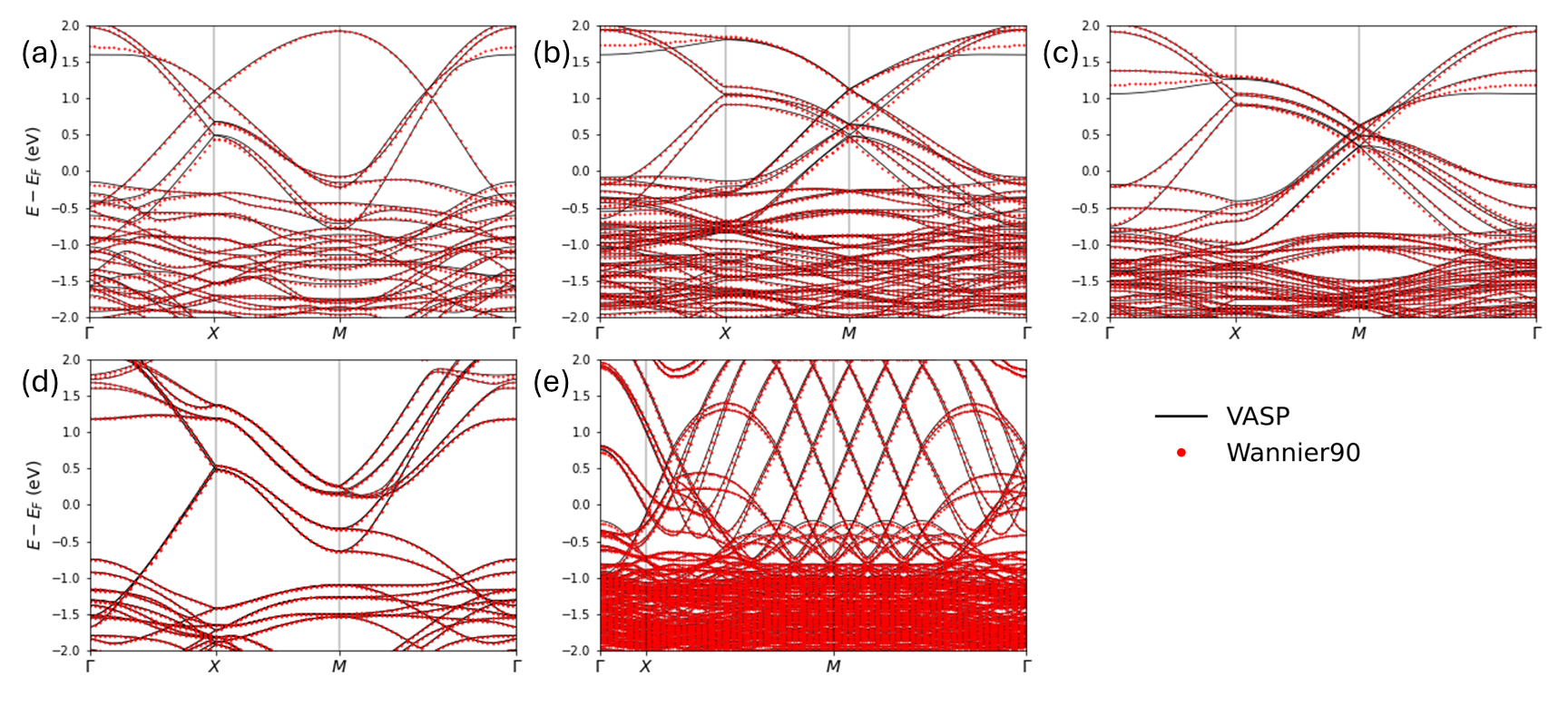}
\end{center}
\caption{
VASP vs Wannier model band structures of (a) YBa$_2$Cu$_3$O$_7$, (b) Pr$_{0.5}$Y$_{0.5}$Ba$_2$Cu$_3$O$_7$, (c) Pr$_{0.5}$YBa$_{1.5}$Cu$_3$O$_7$, (d) Bi$_2$Sr$_2$CaCu$_2$O$_8$, and (e) Bi$_2$Sr$_2$CaCu$_2$O$_{8.25}$.  The $k$-path goes through $\Gamma (0,0,0)$, $X (\pi/a, 0, 0)$, and $M (\pi/a, \pi/b, 0)$ in the reciprocal lattice of the corresponding supercells in Fig.~\ref{fig:crystal}.  Black solid curves and red dots represent the band structures from DFT calculations and Wannier models separately.  
}
\label{fig:wannier}
\end{figure*}
%%%%%%%%%%%%%%%%%%%%%%%%%%%%%%%%%%%%%%%%%%%%%%%%%%%%%

To extract hopping parameters from the DFT calculations, a widely used method is Wannierization, which transforms electronic wavefunctions obtained from plane-wave basis sets into a set of localized real-space functions known as Wannier functions \cite{wannier1937structure}.  By constructing maximally localized Wannier functions \cite{marzari1997maximally}, one can derive tight-binding models from the DFT calculations, using the Wannier basis to provide a more intuitive and physically meaningful interpretation of a material's electronic properties.  This approach is a powerful tool for investigating electronic structure \cite{marzari2012maximally}.  For the Wannierization process, we used the Wannier90 v3.0 package \cite{pizzi2020wannier90}.

Due to the high computational cost, Wannierization is typically performed on a subspace of the full Hilbert space.  The choice of this subspace depends on the orbitals of interest.  In this work, we Wannierized all bands near the Fermi level using Cu $d$-like, O $p$-like, and Bi $p$-like Wannier functions.  This results in a 72-band tight-binding model for YBa$_2$Cu$_3$O$_7$, a 144-band model for Pr$_{0.5}$Y$_{0.5}$Ba$_2$Cu$_3$O$_7$, a 144-band model for Pr$_{0.5}$YBa$_{1.5}$Cu$_3$O$_7$, a 160-band model for  Bi$_2$Sr$_2$CaCu$_2$O$_8$, and a 652-band model for  Bi$_2$Sr$_2$CaCu$_2$O$_{8.25}$.  For computational efficiency, all hoppings smaller than 1 meV were truncated in the final Wannier-derived tight-binding models.

Fig.~\ref{fig:wannier} compares the band structures obtained from VASP with the corresponding Wannierized tight-binding models for each cuprate system studied.  The band structures from both methods are in excellent agreement, demonstrating that the Hilbert subspace spanned by the Cu $d$-like, O $p$-like, and Bi $p$-like orbitals provides an accurate description of the bands near the Fermi level.  This subspace is sufficient to capture both the bonding and anti-bonding bands involved in interlayer band splitting.  Notably, bands far away from the Fermi level with dominant characters not connected to the above Wannier basis (e.g., coming from other atomic orbitals of Ca, Pr, Y, and Ba) can contribute to interlayer processes via weak virtual hybridization processes.  Through Wannierization, these contributions are implicitly included in the hoppings between Wannier functions.

\section{Choice of exchange-correlation functionals}
\label{app:functional}
In this section, we will discuss the effects of different DFT functionals and magnetic orders on the most representative hopping strengths involved in the EICs.  
%%%%%%%%%%%%% TABLE %%%%%%%%%%%%%%%%%%%%%%%%%%%%%%%%%%%
\begin{table*} 
\caption{
The influence of functional choices and magnetic orders on representative hopping strength (in eV) for YBCO7
}
\begin{ruledtabular} 
\begin{tabular} { c c c c c}
Functional & $t_{p_{\sigma}p_{\sigma}}^\perp$ & $t_{p_zp_z}^\perp$ & $t_{d_{x^2}p_z}^\parallel$ \\
\hline
PBE (NM) & 0.201 & 0.574 & 0.265 \\
PBE+$U$ ($U=3$ eV, NM) & 0.206 & 0.564 & 0.266 \\
SCAN (NM) & 0.202 & 0.579 & 0.274 \\
SCAN+$U$ ($U=3$ eV, NM) & 0.203 & 0.582 & 0.274 \\
Maximum relative error (functional choice) & 2.4\% & 3.1\% & 3.3\% \\
PBE+$U$ ($U=3$ eV, GAFM) & 0.207 & 0.567 & 0.261 \\
Relative error (magnetic order) & 0.5\% & 1.2\% & 1.5\% \\

\end{tabular} 
\end{ruledtabular} 
\label{tab:functional}
\end{table*}

%%%%%%%%%%%%%%%%%%%%%%%%%%%%%%%%%%%%%%%%%%%%%%%%%%%%%%%
In Table~\ref{tab:functional}, we tabulate resulting using the PBE, SCAN, PBE+U and SCAN+U functionals for the non-magnetic states with $U=3$ eV applied to the Cu 3d manifold.  The relative errors of hopping strengths induced by different choices functionals are all below 4\% which is acceptable in our use case.  In addition, we tabulate the hopping strengths for the G-AFM magnetic state and compare it with the non-magnetic state with the same DFT functional.  The relative errors of hopping strengths induced by the magnetic order are below 2\% which is also acceptable. 

\section{Effects of long-range in-plane hoppings}
\label{app:ppinpane}
Concerning Fig.~\ref{fig:dppd}, we discussed how the interlayer O $p$-$p$ hoppings facilitate the EICs between Cu $d_{x^2}$ orbitals.  This occurs because the oxygen $p$ orbitals exhibit strong hybridization with the Cu $d_{x^2}$ orbitals via large in-plane hopping processes.  While the primary in-plane $d_{x^2}$-$p$ coupling arises from the direct nearest-neighbor hopping shown in Fig.~\ref{fig:dppd}(c) and (e), several longer-ranged mechanisms also contribute to the in-plane $d_{x^2}$-$p$ coupling.  This section discusses these mechanisms in detail and provides quantitative estimates of their effects. 

%%%%%%%%%%% FIGURE %%%%%%%%%%%%%%%%%%%%%%%%%%%%%%%%%%%
\begin{figure}
\begin{center}
\includegraphics[scale=0.41]{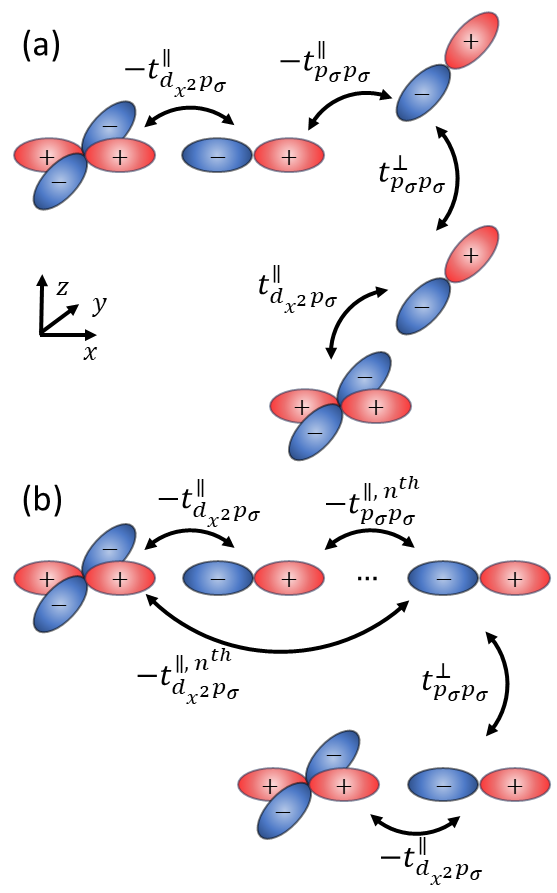}
\end{center}
\caption{
Illustration of EIC mechanisms through interlayer $p_{\sigma}$-$p_{\sigma}$ hopping, where the $d_{x^2}$ orbital and the $p_{\sigma}$ orbital mediating the interlayer coupling from the upper layer are not nearest neighbors.  (a) The $d_{x^2}$ and $p_{\sigma}$ are connected via nearest-neighbor in-plane $p_{\sigma}$-$p_{\sigma}$ and $d_{x^2}$-$p_{\sigma}$ hoppings.  (b) The $d_{x^2}$ and $p_{\sigma}$ are connected via the n$^{th}$ nearest-neighbor in-plane $p_{\sigma}$-$p_{\sigma}$ and nearest-neighbor $d_{x^2}$-$p_{\sigma}$ hoppings, or directly connected via a long-ranged $d_{x^2}$-$p_{\sigma}$ hopping. 
}
\label{fig:apdix_dppd_other}
\end{figure}
%%%%%%%%%%%%%%%%%%%%%%%%%%%%%%%%%%%%%%%%%%%%%%%%%%%%%%

For instance, Fig.~\ref{fig:apdix_dppd_other}(a) illustrates a mechanism where a Cu $d_{x^2}$ orbital is connected to its next-nearest-neighbor O $p_{\sigma}$ orbital in the upper layer via a two-step hopping pathway.  This pathway involves the nearest-neighbor hoppings $t_{d_{x^2}p_{\sigma}}^\parallel$ and $t_{p_{\sigma}p_{\sigma}}^\parallel$. Note that this mechanism is mediated by the interlayer hopping $t_{p_{\sigma}p_{\sigma}}^\perp$, similar to Fig.~\ref{fig:dppd}(c), but features more complex in-plane hopping processes.  The additional $t_{p_{\sigma}p_{\sigma}}^\parallel$ effectively modifies the hybridization strength between the Cu $d_{x^2}$ and O $p_{\sigma}$ orbitals, thereby influencing the corresponding exchange interactions mediated by $t_{p_{\sigma}p_{\sigma}}^\perp$.  An analogous mechanism can be considered by substituting all the O $p_{\sigma}$ orbitals with $p_{z}$ orbitals.

Following the perturbative approach outlined in Sec.~\ref{sec:estimate_EIC}, we incorporate the effects of $t_{p_{\sigma}p_{\sigma}}^\parallel$ and $t_{p_{z}p_{z}}^\parallel$.  These terms improve the accuracy of our predictions for $\Delta_{p_{\sigma}p_{\sigma}\perp}$ and $\Delta_{p_{z}p_{z}\perp}$ by approximately 15 meV for YBCO7.  The corresponding analytical expressions are provided in Appendix~\ref{app:formula}.  

Additionally, Fig.~\ref{fig:apdix_dppd_other}(b) depicts more general in-plane hopping pathways between Cu $d_{x^2}$ and O $p_{\sigma}$ orbitals involving direct long-range hopping or mediation by other oxygen atoms.  In principle, many long-ranged in-plane hoppings can contribute to EICs.  However, these longer-ranged hoppings decay exponentially with distance, and we find that they are at least an order of magnitude smaller than the nearest-neighbor hoppings $t_{d_{x^2}p_{\sigma}}^\parallel$ and $t_{p_{\sigma}p_{\sigma}}^\parallel$ considered here, so are neglected.

\section{Minor mechanisms}
\label{app:mech}
In the main text, we identified the primary EIC mechanisms by sequentially setting $t_{p_{\sigma}p_{\sigma}}^\perp$, $t_{p_zp_z}^\perp$, and $t_{d_{z^2}p_{\sigma}}^\perp$ to zero in the Wannier tight-binding model and tracking the change of the interlayer band splitting.  In addition to these dominant mechanisms, there remain several minor mechanisms that result in negligible EIC contributions and correspond to the residual interlayer band splitting observed in Fig.~\ref{fig:dpdpd}(b).
%\magenta{Recapture that by removing the interlayer mechanisms intermediated by $t_{d_{x^2}d_{x^2}}^\perp$, $t_{p_{\sigma}p_{\sigma}}^\perp$, $t_{p_zp_z}^\perp$, and $t_{d_{z^2}p_{\sigma}}^\perp$, the YBCO7 band structure of the modified tight-binding model becomes Fig.~\ref{fig:dpdpd}(b), where the band splitting caused by EIC almost vanishes.} \red{I don't understand this sentence at all.}  
We quantitatively analyze these weaker EIC mechanisms below.

%%%%%%%%%%% FIGURE %%%%%%%%%%%%%%%%%%%%%%%%%%%%%%%%%%%
\begin{figure*}
\begin{center}
\includegraphics[scale=0.46]{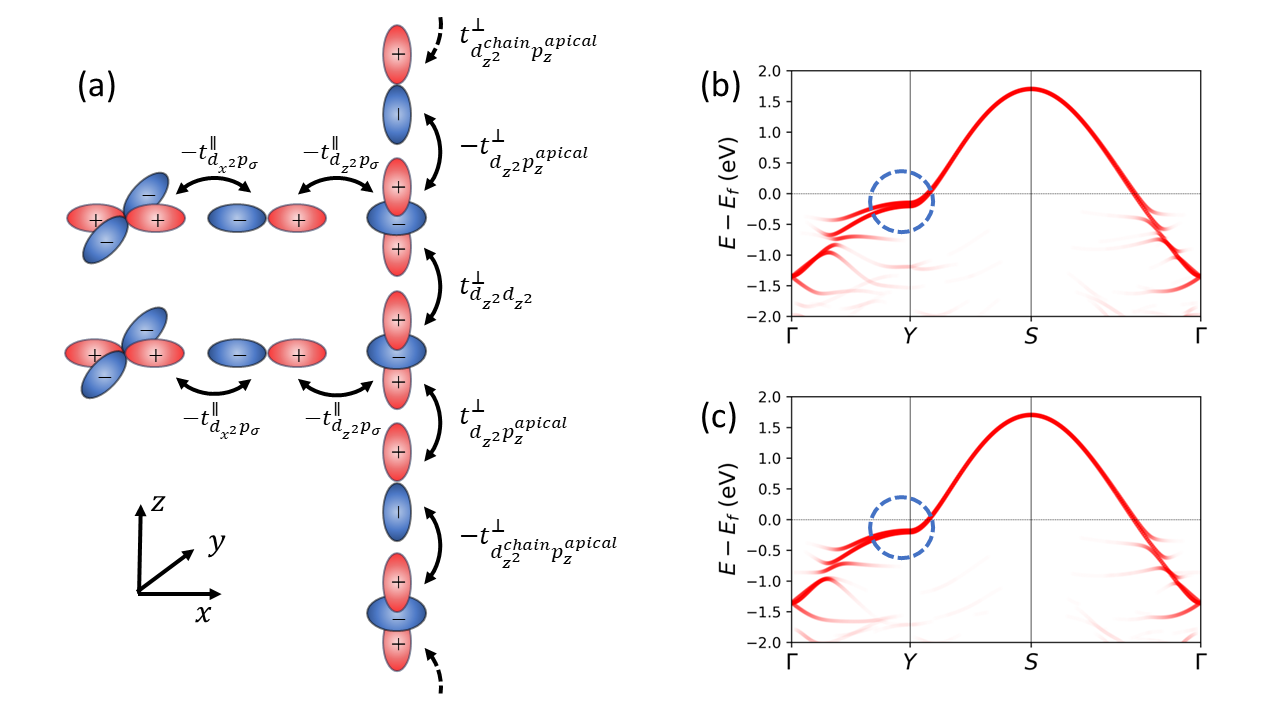}
\end{center}
\caption{
(a) Illustration of an EIC mechanism mediated by interlayer $d_{z^2}$-$d_{z^2}$ hoppings.  In addition, for YBCO specifically, there is an inter-bilayer coupling mechanism intermediated by interlayer $d_{z^2}$-$p_{z}^{\text{apical}}$-$d_{z^2}^{\text{chain}}$-$p_{z}^{\text{apical}}$-$d_{z^2}$ hoppings.  (b) YBCO7 band structure modified from Fig.~\ref{fig:dpdpd}(b) by additionally setting $t_{d_{z^2}d_{z^2}}^\perp=0$.  (c) Band structure modified from (b) by additionally setting $t_{d_{z^2}p_{z}^{\text{apical}}}^\perp=0$. 
}
\label{fig:apdix_dpddpd}
\end{figure*}
%%%%%%%%%%%%%%%%%%%%%%%%%%%%%%%%%%%%%%%%%%%%%%%%%%%%%%

Fig.~\ref{fig:apdix_dpddpd}(a) illustrates two examples of such minor mechanisms.  The first mechanism involves $t_{d_{z^2}d_{z^2}}^\perp$ hopping to create the interlayer connection with the $d_{z^2}$ orbitals acting as a bridge connecting the $d_{x^2}$ orbitals through in-plane $t_{d_{z^2}p_{\sigma}}^\parallel$ and $t_{d_{x^2}p_{\sigma}}^\parallel$ hoppings.  Fig.~\ref{fig:apdix_dpddpd}(b) shows the band structure modified from Fig.~\ref{fig:dpdpd}(b) by setting $t_{d_{z^2}d_{z^2}}^\perp$ to zero.  Compared to Fig.~\ref{fig:dpdpd}(b), the band splitting decreases by about 6 meV at the antinodal region.  This indicates that the contribution to the EIC from this mechanism $\Delta_{d_{z^2}d_{z^2}\perp}^{\text{antinodal}}$ is around 6 meV, significantly smaller than those major mechanisms discussed in the main article.  The band splitting remains unchanged at the nodal region, indicating $\Delta_{d_{z^2}d_{z^2}\perp}^{\text{nodal}}=0$, a symmetry-protected result similar to that of the $\Delta_{d_{z^2}p_{\sigma}\perp}$ mechanism.

The second mechanism works exclusively for the YBCO class of materials which have chain CuO layers between planar CuO$_2$ bilayers.  Distinct from other mechanisms described up to this point, this pathway connects different CuO$_2$ \textit{bilayers} and involves hoppings between Cu $d_{z^2}$ orbitals and apical O $p_{z}$ orbitals along the axial direction.  These hoppings are described by $t_{d_{z^2}p_{z}^{\text{apical}}}^\perp$ and $t_{d_{z^2}^{\text{chain}}p_{z}^{\text{apical}}}^\perp$ as illustrated in Fig.~\ref{fig:apdix_dpddpd}(a).       
% The $d_{z^2}$ orbitals  still play important roles in connecting different layers. \red{They did not before?}  In contrast to the previous mechanism, the EIC between $d_{z^2}$-$d_{z^2}$ orbitals \red{I though EIC was between dx2 only...?  The start of this paragraph is very confusing and needs rewriting} 
%are now given by a chain of hoppings along the $z$-axis: $t_{d_{z^2}p_{z}^{\text{apical}}}^\perp$ and $t_{d_{z^2}^{\text{chain}}p_{z}^{\text{apical}}}^\perp$ hoppings as illustrated in Fig.~\ref{fig:apdix_dpddpd}(a).  
Fig.~\ref{fig:apdix_dpddpd}(c) shows the band structure modified from panel (b) by additionally setting $t_{d_{z^2}^{\text{chain}}p_{z}^{\text{apical}}}^\perp$to zero.  The contribution of inter-bilayer coupling from this mechanism is around 10 meV at the antinodal region and 0 meV at the nodal region.  Since this contribution is significantly smaller than those of the major mechanisms discussed in the main text and is specific to YBCO7, we choose to exclude it from the main article. Nevertheless, investigating materials-specific inter-bilayer coupling mechanisms in various cuprates is an interesting future direction.

\section{In-plane hoppings with low variability}
\label{app:lessvary}
As we have listed in Table~\ref{tab:hops}, some hoppings vary a lot among different materials, while others don't.  One can also notice that those strongly varying hoppings ($t_{\text{max}}/t_{\text{min}}\geq 1.9$) are all related to the out-of-plane ($z$) direction: either interlayer hoppings or coming from $z$-direction distortions.  According to the perturbation formula Eq. \eqref{equ:delta_m}, these strongly varying hoppings are also the most important ones to distinguish different materials.  On the contrary, the hoppings with low variability ($t_{\text{max}}/t_{\text{min}}\leq 1.4$) are all in-plane hoppings, insensitive to the distortions and impurities out of the plane.  These hoppings are relatively unchanged among different cuprates, so one can make a rough estimation for these hoppings without causing large errors when predicting the EIC.  
%%%%%%%%%%% FIGURE %%%%%%%%%%%%%%%%%%%%%%%%%%%%%%%%%%%
\begin{figure*}
\begin{center}
\includegraphics[scale=0.55]{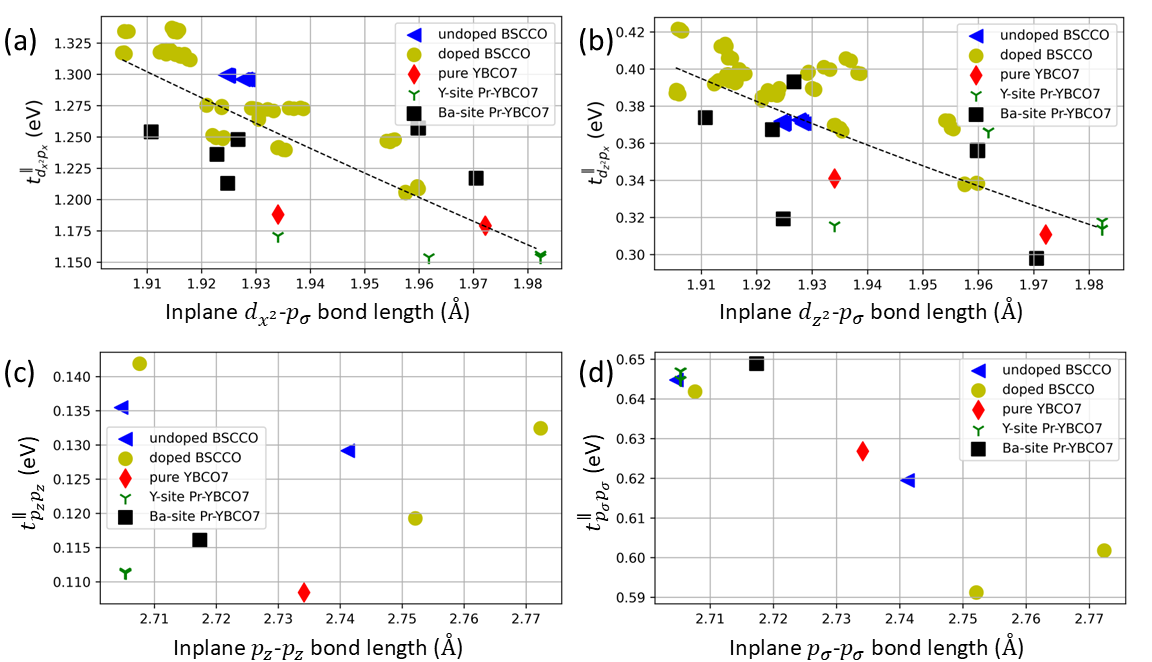}
\end{center}
\caption{
In-plane hoppings contributed to the EIC mechanisms with low variability.  Different data points are from symmetry-independent atoms in undoped BSCCO, 25\% hole-doped BSCCO, YBCO7, Y-site Pr-doped YBCO7, and Ba-site Pr-doped YBCO7. 
(a) In-plane hopping between $d_{x^2}$ and $p_{\sigma}$ orbitals $t_{d_{x^2} p_z}^\parallel$ versus the corresponding in-plane oxygen buckling strength.  The dashed line shows an exponential decay fit.  (b) In-plane hopping between $d_{z^2}$ and $p_{\sigma}$ orbitals.  (c) In-plane hopping between nearest oxygen $p_z$ and oxygen $p_z$ orbitals.  (d) In-plane hopping between nearest oxygen $p_{\sigma}$ and oxygen $p_{\sigma}$ orbitals.
}
\label{fig:apdix_constant_hop_struct}
\end{figure*}
%%%%%%%%%%%%%%%%%%%%%%%%%%%%%%%%%%%%%%%%%%%%%%%%%%%%%%

Fig.~\ref{fig:apdix_constant_hop_struct} shows the strengths of these in-plane hoppings with low variability and their relations with the corresponding bond lengths.  Note that the bond lengths of all these hoppings vary less than 0.1 \AA{} among all the cuprate materials that we have studied, one order of magnitude smaller than the strongly-varying hoppings in Fig.~\ref{fig:hop_struct}.  In practice, when predicting EIC from structural properties, one could either simply approximate these hoppings as constants or use the structure-based formulas given in Appendix~\ref{app:formula}.  Within such a small window of bond length, these two choices make little difference from each other.  More quantitative details are summarized in Appendix~\ref{app:formula}. 

\section{Estimation formula}
\label{app:formula} 
%%%%%%%%%%%%% TABLE %%%%%%%%%%%%%%%%%%%%%%%%%%%%%%%%%%%
\begin{table} 
\caption{
Analytical hopping strength formulas fitted from Fig.~\ref{fig:hop_struct} and Fig.~\ref{fig:apdix_constant_hop_struct}.  The errors are the root-mean-square deviation of predicted values (see text for details). Hoppings and their errors are in eV.  Bond lengths and oxygen buckling are \AA{}.   % \red{I assume bond lengths or buckling are in Angstrom?  And is the error in eV?}
}
\begin{ruledtabular} 
\begin{tabular} { c c c}
Hoppings & Formula & Error \\
\hline
$t_{p_zp_z}^\perp$ (Y, Pr) & $364.460 \exp(-2.273b_{\text{OO}\perp})$ & 0.006 \\ 
$t_{p_{\sigma}p_{\sigma}}^\perp$ & $23.555 \exp(-1.662b_{\text{OO}\perp})$ & 0.006 \\ 
$t_{d_{z^2}p_{\sigma}}^\perp$ & $91.826 \exp(-1.837b_{\text{CuO}\perp})$ & 0.006 \\ 
$t_{d_{x^2}p_z}^\parallel$ & $1.0048b_{\text{buckling}}$ & 0.011 \\
$t_{d_{z^2}p_{\sigma}}^\parallel$ & $168.773 \exp(-3.172b_{\text{CuO}\parallel})$ & 0.023 ($\sim$6\%) \\
$t_{d_{x^2}p_{\sigma}}^\parallel$ & $27.789 \exp(-1.602b_{\text{CuO}\parallel})$ & 0.037 ($\sim$3\%)\\
$t_{p_{\sigma}p_{\sigma}}^\parallel$ & $0.635$ & 0.018 ($\sim$3\%)\\
$t_{p_zp_z}^\parallel$ & $0.119$ & 0.010 ($\sim$8\%)\\
\end{tabular} 
\end{ruledtabular} 
\label{tab:hop_formula}
\end{table}

%%%%%%%%%%%%%%%%%%%%%%%%%%%%%%%%%%%%%%%%%%%%%%%%%%%%%%%
As discussed in Fig.~\ref{fig:hop_struct} and Fig.~\ref{app:lessvary}: the hoppings participating in the EICs are either predicted from structural properties or can be estimated as constants.  In Table~\ref{tab:hop_formula}, we list the analytical formulas of these hoppings and their corresponding errors.  These errors are the root-mean-square deviation of predicted values (dashed lines in Fig.~\ref{fig:hop_struct} and Fig.~\ref{fig:apdix_constant_hop_struct}) from the actual values (data points in the same figures). For the hoppings with low variability, the relative errors compared to the mean hopping strengths are also listed as percentages.  These fitting formulas use structural descriptors to predict hopping strengths.
They are either the interlayer ($\perp$) or intralayer ($\parallel$) bond lengths between two atoms (e.g., $b_{\text{OO}\perp}$ is the bond length in the out-of-plane direction between two nearest neighbor oxygens), or the distortion amplitudes such as oxygen buckling $b_{\text{buckling}}$ in the out-of-plane direction for planar oxygens in the CuO$_2$ layers. 
% \red{How did you get the standard error? I mean what is is compared to?}. 
% \red{Define the buckling}

We choose exponential decaying forms of the bond lengths for the formulas due to the localized nature of Wannier functions.  One could also approximate these formulae with linear relations for most cases shown in Fig.~\ref{fig:hop_struct} and Fig.~\ref{app:lessvary} due to the relatively narrow range of bond lengths.  Note that $t_{p_zp_z}^\perp$ has two different formulae depending on the type of atomic layer (such as Ca, Y, and Pr) sandwiched by the CuO$_2$ bilayers.  Since we project out the high-energy empty orbitals of these atoms as well as their (semi-)core states during the Wannierization, $t_{p_zp_z}^\perp$ includes any effective interlayer hopping between $p_z$ orbitals mediated by them.  Hence, the type of sandwiched atom can affect $t_{p_zp_z}^\perp$.  We provide the $t_{p_zp_z}^\perp$ formulae for BSCCO (with Ca atoms) and YBCO (with Y or Pr atoms) separately.  Understanding the detailed hopping mechanisms that involve the sandwiched atoms is beyond the scope of current work but may be worth studying in the future.  For simplicity, the averaged hopping strengths for each material will be used to estimate the EIC $\Delta_M$ for each mechanism.  This simplification increases the errors of hopping estimate removes the potential need to identify all atomic positions in experiments or DFT relaxations.  

In addition to the hopping strength values, the off-diagonal matrix elements of the tight-binding Hamiltonian are also related to the reciprocal lattice point $k$ and thus the phases of the hopping processes.  For example, for a minimum model where the unit cell consists of two CuO$_2$ layers with one Cu per layer, $t_{d_{x^2}p_{\sigma}}^\parallel$ describes the hopping strength from $d_{x^2}$ orbital to $p_{\sigma}$ orbital.  One would further notice that this hopping is negative towards $+x$ direction, while the one towards $-x$ is positive.  Therefore, the off-diagonal matrix element hopping from the $d_{x^2}$ orbital to the nearest $p_{\sigma}$ orbital along the $x$-axis is 
$$
t_{d_{x^2}p_{\sigma}}^\parallel\left(-e^{-ik_xa/2}+e^{ik_xa/2}\right)
=2i t_{d_{x^2}p_{\sigma}}^\parallel\sin\left(k_xa/2\right)
$$
where $a$ is the lattice constant along $x$ axis (nearest in-plane Cu-Cu distance), $k_x$ is the $x$-component of the reciprocal vector $k$.  The reversed hopping from the $p_{\sigma}$ orbital to the nearest $d_{x^2}$ orbital is $-2i t_{d_{x^2}p_{\sigma}}^\parallel\sin\left(k_xa/2\right)$.  On the contrary, for the hopping towards $\pm y$ direction, the $d_{x^2}$ orbitals display negative wavefunction along the $y$-axis, so all the signs of the corresponding matrix elements should be switched compared to the $x$-axis hoppings. 

Some other hopping procedures are not dependent on $k_x$ and $k_y$.  For example, $t_{d_{x^2}d_{x^2}}^\perp$, $t_{p_{\sigma}p_{\sigma}}^\perp$ and $t_{p_zp_z}^\perp$ are hoppings towards $z$-direction, so the corresponding off-diagonal matrix elements are proportional to $e^{ik_zh}$, where $h$ is the corresponding inter-layer bond-length.  Detailed $k$-dependency of different hopping processes are listed in Table~\ref{tab:kdep_hopping}.  

%%%%%%%%%%%%% TABLE %%%%%%%%%%%%%%%%%%%%%%%%%%%%%%%%%%%
\begin{table} 
\caption{
The $k$-dependency of different hopping processes.  Recall that we have defined $p_{\sigma}$ as the oxygen $p$-orbital pointing towards nearest-neighbor Cu atoms in this article.
}
\begin{ruledtabular} 
\begin{tabular} { c c}
Hopping & $k$-dependency \\
\hline
$t_{d_{x^2}d_{x^2}}^\perp$, $t_{p_{\sigma}p_{\sigma}}^\perp$, $t_{p_zp_z}^\perp$ & $e^{ik_zh}$ \\ 
$t_{d_{x^2}p_{\sigma}}^\parallel$ (along $x$) & $2i \sin\left(k_xa/2\right)$ \\
$t_{d_{x^2}p_{\sigma}}^\parallel$ (along $y$) & $-2i \sin\left(k_ya/2\right)$ \\
$t_{d_{x^2}p_z}^\parallel$ (along $x$) & $2 \cos\left(k_xa/2\right)$ \\
$t_{d_{x^2}p_z}^\parallel$ (along $y$) & $-2 \cos\left(k_ya/2\right)$ \\
$t_{p_{\sigma}p_{\sigma}}^\parallel$ & $4 \sin\left(k_xa/2\right) \sin\left(k_ya/2\right)$\\
$t_{p_{\sigma}d_{z^2}}^\parallel$ (along $x$ or $y$) & $2i \sin\left(k_{x(y)}a/2\right)$ \\
$t_{d_{z^2}p_{\sigma}}^\perp$ (along $x$ or $y$) & $-2i \sin\left(k_{x(y)}a/2\right) e^{ik_zh}$ \\
\end{tabular} 
\end{ruledtabular} 
\label{tab:kdep_hopping}
\end{table}
%%%%%%%%%%%%%%%%%%%%%%%%%%%%%%%%%%%%%%%%%%%%%%%%%%%%%%%

%%%%%%%%%%%%% TABLE %%%%%%%%%%%%%%%%%%%%%%%%%%%%%%%%%%%
\begin{table} 
\caption{
Analytically derived $k$-dependency of different mechanisms $D_M(k)$
}
\begin{ruledtabular} 
\begin{tabular} { c c}
Mechanism & $k$-dependency \\
\hline
$t_{d_{x^2}p_{\sigma}}^\parallel t_{p_{\sigma}p_{\sigma}}^\perp t_{p_{\sigma}d_{x^2}}^\parallel$ & $\sin^2\frac{k_xa}{2}+\sin^2\frac{k_ya}{2}$ \\
$t_{d_{x^2}p_z}^\parallel t_{p_zp_z}^\perp t_{p_zd_{x^2}}^\parallel$ & $\cos^2\frac{k_xa}{2}+\cos^2\frac{k_ya}{2}$ \\
$t_{d_{x^2}p_{\sigma}}^\parallel t_{p_{\sigma}p_{\sigma}}^\parallel t_{p_{\sigma}p_{\sigma}}^\perp t_{p_{\sigma}d_{x^2}}^\parallel$ & $-\sin^2\frac{k_xa}{2}\sin^2\frac{k_ya}{2}$ \\
$t_{d_{x^2}p_z}^\parallel t_{p_zp_z}^\parallel t_{p_zp_z}^\perp t_{p_zd_{x^2}}^\parallel$ & $\cos^2\frac{k_xa}{2}\cos^2\frac{k_ya}{2}$ \\
$t_{d_{x^2}p_{\sigma}}^\parallel t_{p_{\sigma}d_{z^2}}^\parallel t_{d_{z^2}p_{\sigma}}^\perp t_{p_{\sigma}d_{x^2}}^\parallel$ & $\left(\sin^2\frac{k_xa}{2}-\sin^2\frac{k_ya}{2}\right)^2$ \\
\end{tabular} 
\end{ruledtabular} 
\label{tab:kdep}
\end{table}
%%%%%%%%%%%%%%%%%%%%%%%%%%%%%%%%%%%%%%%%%%%%%%%%%%%%%%%

Armed with the hopping strengths, we derive the $k$-dependency of each major EIC mechanism following the procedure discussed in Sec.~\ref{sec:estimate_EIC} and list them in Table~\ref{tab:kdep}.  Each mechanism is identified by the chain of hoppings that connect the $d_{x^2}$ orbitals from different layers.  
% \magenta{We have not carefully investigated inter-bilayer coupling mechanisms involving inter-bilayer couplings: depending on the inter-bilayer atoms, such mechanisms are material-specific and require further studies.} \red{what are you talking about here? the sandwiched stuff?  or something else?  very confusing.}  As a simplified intra-bilayer picture, the mechanisms described in this work do not depend on the $k_z$ component as expected.
% These mechanisms usually involve multiple paths of hoppings along $x$- or $y$-axis, or a combination of them.  For example, the mechanism $t_{d_{x^2}p_{\sigma}}^\parallel t_{p_{\sigma}p_{\sigma}}^\perp t_{p_{\sigma}d_{x^2}}^\parallel$ involves a path along $x$-axis, showing the $2i \sin\left(k_xa/2\right)*(-2i \sin\left(k_xa/2\right))=4\sin^2\frac{k_xa}{2}$ dependency on the reciprocal lattice vector.  In addition, this mechanism also involves a path along $y$-axis.  For simplicity, we are assuming tetragonal symmetry so the total $k$-dependency is $\sin^2\frac{k_xa}{2}+\sin^2\frac{k_ya}{2}$.  \red{this paragraph seems redundant given the main text and your prior text in this section; please remove or clean up.}

It is interesting that the fitting formula Eq. \eqref{equ:old} used in prior literature indeed captures the $k$-dependency of the $t_{d_{x^2}d_{x^2}}^\perp$ and  $t_{d_{x^2}p_{\sigma}}^\parallel t_{p_{\sigma}d_{z^2}}^\parallel t_{d_{z^2}p_{\sigma}}^\perp t_{p_{\sigma}d_{x^2}}^\parallel$ mechanisms because it can be rewritten as 
$$
\frac{1}{4}\left[\text{cos}(k_xa) – \text{cos}(k_ya)\right]^2 =\left(\sin^2\frac{k_xa}{2}-\sin^2\frac{k_ya}{2}\right)^2\,.
$$
However, the mechanisms mediated by $t_{p_{\sigma}p_{\sigma}}^\perp$ or $t_{p_zp_z}^\perp$ are missing completely in Eq. \eqref{equ:old} as they have a different analytical form.

We summarize the perturbative estimation formulae of the major EIC mechanisms.  The total EIC strength is the sum of the following mechanisms:
\begin{multline*}
\Delta_{d_{z^2}p_{\sigma}\perp}=0.9116 \bar t_{d_{x^2}p_{\sigma}}^{\parallel 2} \bar t_{p_{\sigma}d_{z^2}}^\parallel \bar t_{d_{z^2}p_{\sigma}}^\perp \\\left(\sin^2\frac{k_xa}{2}-\sin^2\frac{k_ya}{2}\right)^2\,,
\end{multline*}
\begin{multline*}
\Delta_{p_{\sigma}p_{\sigma}\perp}=\bar t_{d_{x^2}p_{\sigma}}^{\parallel 2} \bar t_{p_{\sigma}p_{\sigma}}^\perp \left[0.4495\left(\sin^2\frac{k_xa}{2}+\sin^2\frac{k_ya}{2}\right)\right.\\
\left.-1.428\bar t_{p_{\sigma}p_{\sigma}}^{\parallel}\sin^2\frac{k_xa}{2}\sin^2\frac{k_ya}{2}\right]\,,
\end{multline*}
\begin{multline*}
\Delta_{p_zp_z\perp}=\bar t_{d_{x^2}p_z}^{\parallel 2} \bar t_{p_zp_z}^\perp \left[0.3469\left(\cos^2\frac{k_xa}{2}+\cos^2\frac{k_ya}{2}\right)\right.\\
\left.+21.59\bar t_{p_zp_z}^{\parallel}\cos^2\frac{k_xa}{2}\cos^2\frac{k_ya}{2}\right]\,.
\end{multline*}
The energy units are eV in the above expressions.  The average hopping strengths can be predicted from structural properties as listed in Table~\ref{tab:hop_formula}.  The coefficients in these formulae contain more detailed information beyond hopping strengths such as onsite energy differences, which are approximately invariant among different cuprates and defined by fitting the five cuprates materials on their nodal and antinodal points as discussed in the main text.  The hoppings $\bar t$ in these formulae can be predicted by averaged structural properties following Table \ref{tab:hop_formula}.  An example usage of these formulae has been discussed in Sec.~\ref{sec:example}, where we extended the prediction of EICs in YBCO7 onto the whole BZ. 

\bibliography{v1}
\end{document}